\documentclass [twoside,10pt] {article}
\usepackage{setspace,graphicx,bm,fancyhdr}
\usepackage{amsmath,amsthm,amssymb,latexsym}
\usepackage{color}
\usepackage{graphicx}

\usepackage{authblk}

\newcommand{\vct}[1]{\mbox{$\stackrel{\rightarrow}{#1}$}}

\newcommand{\slas}[1]{\mbox{${{#1} \!\!\! /}$}}

\begin{document}
\title{\bf Computation of the masses of the elementary particles} 
\author{John Mashford \\
School of Mathematics and Statistics \\
University of Melbourne, Victoria 3010, Australia \\
E-mail: mashford@unimelb.edu.au\\
ORCID: https://orcid.org/0000-0001-6100-031X}
\date{\today}
\maketitle

\begin{abstract}
An approach to gauge theory in the context of locally conformally flat space-time is described. It is discussed how there are a number of natural principal bundles associated with any given locally conformally flat space-time $X$. The simplest of these principal bundles is the bundle $P_X(G)$ with structure group $G=U(2,2)$. An 11 dimensional bundle $Q$ with structure group a certain 7 dimensional group $K$ is constructed by a method involving a reduction of structure group for the bundle $P_X(G)$.  It is shown how the gauge groups $U(1)$, $SU(2)$ and $SU(3)$ can be derived from the geometry of locally conformally flat space-time. Fock spaces of multiparticle states for the fields of the standard model are constructed in the context of bundles with these groups as structure groups. Scattering and other particle interaction processes are defined in terms of linear maps between multiparticle state spaces. A technique for computing analytically and/or computationally the masses of the elementary particles is described. This method involves the computation of a certain quantity called the integral mass spectrum for a given family of particles and then the masses of the particles in the family are determined to be the locations of the peaks of the integral mass spectrum. The method is applied successfully in the electroweak sector to the cases of the charged leptons, $\mu$ and $\tau$, and the Z$^0$ particle.
\end{abstract}


\section{Introduction}

This paper is a preprint for a paper \cite{AIPAdv} which was previously published before the announcement of this preprint. The published paper had some technical issues which are completely resolved in this preprint and a forthcoming preprint.
 
The standard model (SM) has been described as the most successful physical theory ever conceived. It is one of the twin pillars of modern physics along with general relativity (GR). The SM takes place in a flat space-time arena. The fundamental outputs of the SM are the Feynman amplitudes. Through these amplitudes the theory has been compared to experiment with precise agreement.

The Feynman amplitudes are objects with (polarization, and other) indices and continuous arguments for which the continuous arguments range over Minkowski space ${\bf R}^4$ (or, more specifically, over the ``mass shell" hyperboloids $H_m\subset{\bf R}^4$).

We model space-time as a locally conformally flat Lorentzian manifold $X$, i.e. a particular type of GR space-time \cite{AMP}. In our work the flat arena over which the continuous Feynman amplitude arguments range is the tangent space $T_xX$ at any given point $x\in X$.

The aim of the present paper is to describe a mathematical framework and then to consider the problem of determining the masses of the elementary particles.

The main result of the paper is the description of a method for computing the masses of the elementary particles.  In contrast to other approaches described in the literature our method only involves simple tree level computations.

\section{A brief review of previous work concerning calculation of the masses of the elementary particles}

The main approaches to determining the masses of the elementary particles which have been presented in the literature are
\begin{itemize}
\item generation of higher generation masses by radiative corrections of masses associated with the tree level
\item horizontal symmetry
\item symmetries associated with discrete subgroups of gauge groups
\item approximate formulae such as the Fritzsch form ansatz
\item seesaw mechanism
\end{itemize}

The idea that particle masses arise from radiative corrections has a long history dating back to Weinberg \cite{Weinberg} and Georgi and Glashow \cite{Georgi}. It is motivated by the observation that $m_e/m_{\mu}=O(\alpha)$.

Balakrishna {\em et al.} \cite{Balakrishna} describe an approach whereby fermion masses may follow from radiative corrections with no horizontal symmetry required. They postulate the existence of a softly broken discrete charge symmetry. They only consider the case of quarks. Their approach works roughly up to a factor of about 2.

Mohanta and Patel \cite{Mohanta} discuss an approach to elementary particle mass generation where the third generation particles become massive at leading order while the masses of the first two generations arise from quantum (radiative) corrections. Their model has a large number of parameters which makes it less predictive.

Dobrescu and Fox \cite{Dobrescu} propose a mechanism for generating quark and lepton masses based on the assumption that only the top quark mass arises at tree level. New fields are required to couple the top quark to SM fermions. There are 24 parameters in their model.

Barger {\em et al.} \cite{Barger} consider an approach in the context of a fourth generation of quarks.

Horizontal symmetry is a proposed intergenerational symmetry based on discrete or continuous groups. Many horizontal symmetry groups have been proposed in the literature \cite{Lam} including $Z_m,Z_m\times Z_n,D_n,SO(3)$ and $SU(3)$. Leurer {\em et al.} \cite{Leurer} obtain some results concerning horizontal symmetry and spontaneous symmetry breaking.

Barbieri {\em et al.} \cite{Barbieri} consider the relation between the Cabibbo angle and the quark masses in terms of discrete symmetries of the $SU(2)\times U(1)$ model.

Frampton and Klephart \cite{Frampton} discuss an approach to generating particle masses using discrete flavor groups $G$ embedded in $SU(2)$. The top quark becomes heavy because it has a $G$ invariant mass. All the other quarks are considered to break $G$ invariance. The scheme is qualititave but not quantitative.

Barr \cite{Barr} proposes a model based on the group $E_6\times Z_2$. However additional fields are required for a more realistic model than the simple form dictated by the $E_6\times Z_2$ symmetry.

The so called Fritzsch form is an ansatz leading to a relation between the KM angles and the ratios between the fermion masses \cite{Fritzsch}. There are no strong arguments for the Fritzsch form or other ``forms" proposed in the literature \cite{Barr}. 

The seesaw mechanism \cite{Davidson} is an approach to trying to understand the very low masses of the neutrinos compared to the masses of the other elementary particles.

Georgi and Glashow \cite{Georgi} describe some of the approximate relations that are satisfied by the elementary particle masses such as the Gell-Mann-Okubu mass formula (which has now been superceded by QCD considerations).

Froggatt {\em et al.} \cite{Froggatt} argue that the SM should be considered to be a low energy remnant of a larger group and that the observed fermion masses are the result of symmetry breaking in this larger model.

Komatsu \cite{Komatsu} argues that in certain grand unified theories one may obtain bounds on particle masses in the case where there would be four generations of fermions.

In general, while considerable work has been put into the problem of computing the masses of the elementary particles it is fair to say that the problem has not yet been satisfactorily resolved.

\section{Description of the mathematical framework}
 
\subsection{Space-time geometry, $U(2,2)$ covariance and $K$ covariance}

We model space-time as a causal structure which is locally isomorphic to the Minkowski space causal structure  which, in the case of locally Euclidean topology, is equivalent to modelng it as a (causal) 4D Lorentzian M\"{o}bius structure $X=(X,{\mathcal A})$ where ${\mathcal A}=\{\phi_i:i\in I\}$ is the maximal atlas of charts for $X$ (see Ref.~\cite{AMP}). This is equivalent to modeling it as a  locally conformally flat 4D Lorentzian manifold. 

Let $G=U(2,2)$ where the group $U(2,2)$ is taken to be with respect to the hermitian form
\[ (u,v)=u^{\dagger} gv,\forall u,v\in{\bf C}^{4}, \]
where $g$ is the matrix 
\[ g=\gamma^0=\left(\begin{array}{cc}
0&1\\
1&0
\end{array}\right)(=\left(\begin{array}{cc}
0&1_2\\
1_2&0
\end{array}\right)). \]
In our work we take, unless otherwise stated, the Dirac gamma matrices $\gamma^{\mu}$ to be in the Weyl (chiral) representation.

Thus
\[ G=\left\{A\in GL(4,{\bf C}):A^{\dagger}gA=g\right\}. \]

Let $H\subset G$ be the group
\[ H=\left\{\left(\begin{array}{cc}
a&b\\
c&d
\end{array}\right)\in U(2,2):b=0\right\}. \]

There is associated with $X$ a natural smooth principal fiber bundle \cite{Kobayashi} $P_X(G)$ with structure group $G$ and a natural reduction of structure group to a smooth principal bundle $P_X(H)$ with structure group $H$ (see Ref.~\cite{AMP}). 

Letting $K\subset H$ be the group
\[ K=\left\{\left(\begin{array}{cc}
a&0\\
0&a^{\dagger-1}\end{array}\right):a\in GL(2,{\bf C}),|\mbox{det}(a)|=1\right\}, \]
one can obtain \cite{AMP} a smooth principal fiber bundle $Q=P_X(K)$ with structure group $K$ as a homomorphic image of $P_X(H)$, where $Q$ has dimension $\mbox{dim}(Q)=11$.

\subsection{An approach to derivation of the gauge groups $U(1),SU(2)$ and $SU(3)$ from space-time geometry\label{section:gauge_groups}} 

In general, for a princpal bundle $P_X(G)$ with structure group $G$ and transition functions $g_{ij}$, elements of the fiber $\pi^{-1}(x)\subset P_X(G)$ where $x\in X$ and $\pi:P_X(G)\rightarrow X$ is the standard projection, can be thought of as maps $(i\mapsto g_i):I_x\rightarrow G$ such that
\begin{equation}\label{eq:bundle_condition}
g_i=g_{ij}(x)g_j,\forall i,j\in I_x, 
\end{equation}
where, for $x\in X$, $I_x=\{i\in I:x\in\mbox{domain}(\phi_i)\}$ is the collection of indices of charts about $x$. 

Now consider the case when $X$ is a 4D M\"{o}bius structure $X=(X,{\mathcal A})$.
For any $x\in X$ let $I_{0x}\subset I$ denote the set of indices of charts about $x$ where two charts are identified if they agree on a neighborhood of $x$. Thus elements of $I_{0x}$ correspond to equivalence classes of charts, which may be described as germs of charts. 

There is, for any fixed reference chart index $r\in I_{0x}$, by virtue of Liouville's theorem concerning conformal transformations in ${\bf R}^n$ for $n>2$, a one to one correspondence between $I_{0x}$ and $G$. Thus elements of $\pi^{-1}(x)$ can be thought of as maps $\Phi:G\rightarrow G$ which satisfy
\begin{equation}\label{eq:bundle_condition_1}
\Phi(g)=\Gamma_x(g,h)\Phi(h),\forall g,h\in G,
\end{equation} 
where $\Gamma_x:G\times G\rightarrow G$.

\subsection{Infinitesimal transition functions and the ${\mathfrak g}=u(2,2)$ Lie algebra bundle $P({\mathfrak g})$}

For any $x\in X$ we may  differentiate Eq.~\ref{eq:bundle_condition_1} at $(e,e)$, where $e=1_4$, to obtain
\[ ((D\Phi)(e))(A)=((D\Gamma_x)(e,e))(A,B)\Phi(e)+\Gamma_x(e,e)((D\Phi)(e))(B),\forall A,B\in{\mathfrak g}, \]
where ${\mathfrak g}=u(2,2)$ is the Lie algebra of $G$. 
But $\Phi(e)=e$ and $\Gamma_x(e,e)=e$. Therefore writing $((D\Phi)(e)))(A)=\Phi(A)$ and $((D\Gamma_x)(e,e))(A,B) =\Gamma_x(A,B)$ for all $A,B\in{\mathfrak g},x\in X$, we have, at the infinitesimal level, the set $P_x({\mathfrak g})$ defined to be the set of all maps $\Phi:{\mathfrak g}\rightarrow{\mathfrak g}$ which satisfy 
\begin{equation}\nonumber
\Phi(A)=\Gamma_x(A,B)+\Phi(B),\forall A,B\in{\mathfrak g}.
\end{equation}
We call $(x,A,B)\mapsto\Gamma_x(A,B)$ the infinitesimal transition functions associated with the transition functions $g_{ij}$ for $P_X(G)$. We have $\Gamma_x:{\mathfrak g}\times{\mathfrak g}\rightarrow{\mathfrak g}$ and $\Gamma_x$ satisfies the infinitesimal cocycle condition
\begin{equation}
\Gamma_x(A,C)=\Gamma_x(A,B)+\Gamma_x(B,C),\forall A,B,C\in{\mathfrak g},x\in X.
\end{equation}

The bundle $P_X(G)$, as specified by the transition functions $g_{ij}$, is made up of a collection of fibers $\pi^{-1}(x)$ for $x\in X$ for which each fiber is diffeomorphic to $G$. In an analogous fashion the infinitesimal transition functions are associated with the bundle $P({\mathfrak g})=\bigcup_{x\in X}P_x({\mathfrak g})$ (disjoint union) of Lie algebras over $X$.
Any two elements of $P_x({\mathfrak g})$ differ by a constant since for all $\Phi_1,\Phi_2\in P_x({\mathfrak g})$
\[ (\Phi_1-\Phi_2)(A)=(\Gamma_x(A,B)+\Phi_1(B))-(\Gamma_x(A,B)+\Phi_2(B))=(\Phi_1-\Phi_2)(B),\forall A,B\in{\mathfrak g}, \]
and hence $\Phi_2=\Phi_1+C$ for some $C\in{\mathfrak g}$. Conversely, given $\Phi_1\in P_x({\mathfrak g})$ and $C\in{\mathfrak g}$ and defining $\Phi_2(A)=\Phi_1(A)+C$ for all $A\in{\mathfrak g}$ we have
\[ \Phi_2(A)=\Phi_1(A)+C=(\Gamma_x(A,B)+\Phi_1(B))+C=\Gamma_x(A,B)+(\Phi_1(B)+C)=\Gamma_x(A,B)+\Phi_2(B), \]
$\forall A,B\in{\mathfrak g}$, so $\Phi_2\in P_x({\mathfrak g})$. Therefore $P_x({\mathfrak g})$ is an affine space which may be identified with ${\mathfrak g}$.

Now recall how given a principal bundle $P=P_X(G)$ defined by transition functions $g_{ij}$ we have the natural right action $(p,g)\in P\times G\mapsto pg\in P$ defined by $(pg)_i=p_ig$ and we can verify that for all $x\in X,p\in\pi^{-1}(x),g\in G$
\[ (pg)_i=p_ig=(g_{ij}(x)p_j)g=g_{\ij}(x)(p_jg)=g_{ij}(x)(pg)_j,\forall i,j\in I_x, \]
which proves that $pg\in\pi^{-1}(x)\subset P$.

We have by analogy with the operation of right multiplication of $G$ on the principal bundle $P_X(G)$ an operation of right ``multiplication" of ${\mathfrak g}$ on $P({\mathfrak g})=\bigcup_{x\in X}P_x({\mathfrak g})$ given by $(\Phi,A)\in P_x({\mathfrak g})\times{\mathfrak g}\mapsto\Phi A$ where
\begin{align*}
(\Phi A)(B)&=\Phi(B)+A,
\end{align*}
Then we can check that
\begin{align*}
(\Phi A)(B)&=\Phi(B)+A=(\Gamma_x(B,C)+\Phi(C))+A=\Gamma_x(B,C)+(\Phi(C)+A)\\
&=\Gamma_x(B,C)+(\Phi A)(C),\forall A,B,C\in{\mathfrak g},
\end{align*}
which proves that $\Phi A\in P_x({\mathfrak g})$. Therefore right multiplication is a map from $P_x({\mathfrak g})\times{\mathfrak g}$ to $P_x({\mathfrak g})$.

\subsection{The linear mappings from ${\mathfrak g}$ to $u(2)\cong(u(1)\times su(2))$ and $su(3)$ and the corresponding infinitesimal transition functions}

Let \[ A=\left(\begin{array}{cc}
a&b\\
c&d
\end{array}\right)\in gl(4,{\bf C})={\bf C}^{4\times4}\text{ where }a,b,c,d\in{\bf C}^{2\times2}.\]
Then
\begin{align*} 
A\in u(2,2)&\Leftrightarrow A^{\dagger}=-gAg\\
&\Leftrightarrow\left(\begin{array}{cc}
a^{\dagger}&c^{\dagger}\\
b^{\dagger}&d^{\dagger}
\end{array}\right)=-\left(\begin{array}{cc}
0&1\\
1&0
\end{array}\right)\left(\begin{array}{cc}
a&b\\
c&d
\end{array}\right)\left(\begin{array}{cc}
0&1\\
1&0
\end{array}\right)=\left(\begin{array}{cc}
-d&-c\\
-b&-a
\end{array}\right)\\
&\Leftrightarrow c^{\dagger}=-c,b^{\dagger}=-b,d=-a^{\dagger}. \end{align*}
Thus we have natural linear maps $\Xi_1:{\mathfrak g}\rightarrow u(2)$ and $\Xi_2:{\mathfrak g}\rightarrow gl(2,{\bf C})$ defined by
\[ \Xi_1(A)=b, \]
and
\[ \Xi_2(A)=a. \]

The mapping $\Xi_1$ has the property that it is bijective on $S_1=\left\{\left(\begin{array}{cc}
0&b\\
0&0
\end{array}\right):b\in u(2)\right\}\subset{\mathfrak g}$ and $\Xi_2$ has the property that it is bijective on $S_2=\left\{\left(\begin{array}{cc}
a&0\\
0&a^{\dagger}
\end{array}\right):a\in gl(2,{\bf C}\right\}\subset{\mathfrak g}$.

Viewing $gl(2,{\bf C})$ as a real Lie algebra we have that $|gl(2,{\bf C})|=|su(3)|=8$. Let $\Theta$ be any real vector space isomorphism from $gl(2,{\bf C})$ to $su(3)$.
Then there are maps 
\begin{align*}
\Gamma_{1,x}(A,B)&=\Xi_1(\Gamma_x(A,B))\in u(2),\\
\Gamma_{2,x}(A,B)&=\Theta(\Xi_2(\Gamma_x(A,B)))\in su(3),
\end{align*} 
for $x\in X,A,B\in{\mathfrak g}$. The maps $\Gamma_{1}:X\times{\mathfrak g}\times{\mathfrak g}\rightarrow u(2)\mbox{ and }\Gamma_{2}:X\times{\mathfrak g}\times{\mathfrak g}\rightarrow su(3)$ may be considered to be infinitesimal $u(2)\cong(u(1)\times su(2))$ and $su(3)$ valued transition functions.

Now define $U_{2,x}(A,B)=\exp(\Gamma_{2,x}(A,B))$ for all $A,B\in{\mathfrak g}$. Then $U_{2,x}(A,B)\in SU(3),\forall x\in X,A,B\in{\mathfrak g}$.
Similarly we can exponentiate the values taken by $\Gamma_{1,x}$ to obtain $U(2)$ valued functions, noting that $U(2)$ is locally isomorphic to $U(1)\times SU(2)$.

\section{The electroweak Fock spaces resulting from $U(2)$ gauge invariance\label{section:EW_Fock_space}}

Identifying ${\bf R}^4$ in the usual way with $iu(2)=\{\mbox{ hermitian $2\times2$ matrices }\}$ according to $p\mapsto M(p)=p^{\mu}\sigma_{\mu}$ where $\{\sigma_{\mu}\}_{\mu=1}^3$ are the Pauli matrices and $\sigma_0=1_2$, $U(2)$ acts on ${\bf R}^4$ according to
\[ ap=a(M(p))=aM(p)a^{\dagger}=aM(p)a^{-1}. \]
We define the Fock space ${\mathcal F}_{k_1,k_2,k_3,k_4}$ of tensor valued functions $u:\{0,1,2,3\}^{k_1}\times\{1,2\}^{k_2}\times\{0,1,2,3\}^{k_3}\times({\bf R}^4)^{k_4}\rightarrow{\bf C}$ where $k_1,k_2,k_3,k_4\in\{0,1,2,\ldots\}$ and ${\bf C}$ denotes the complex numbers.

If $u^{\mu\alpha i}\in{\mathcal F}_{k_1,k_2,k_3,k_4}$ then $\mu=(\mu_1,\ldots,\mu_{k_1})$ are Lorentz indices, $\alpha=(\alpha_1,\ldots,\alpha_{k_2})$ are polarization indices, $i=(i_1,\ldots,i_{k_3})$ are $u(2)$ indices and $u^{\mu\alpha i}:({\bf R}^4)^{k_4}\rightarrow{\bf C}$. 

We require that $u\in{\mathcal F}_{k_1,k_2,k_3,k_4}$ be Schwartz \cite{Friedlander,IJMPA} in its continuous arguments and relax that condition to obtain the space ${\mathcal P}_{k_1,k_2,k_3,k_4}$ of complex tensor valued functions $u$ which are smooth and polynomially bounded in their continuous arguments. 

When $u(2)$ indices are 0 they correspond to the Lie algebra $u(1)$, hence the group $U(1)$, i.e. QED. When they are in $\{1,2,3\}$ they correspond to the Lie algebra $su(2)$, hence the group $SU(2)$, i.e. the weak interaction.

The Fock space ${\mathcal F}_{k_1,k_2,k_3,k_4}$ is considered to be a space of multiparticle states with $k_4$ particles.

The Fock spaces ${\mathcal F}_{0,k_2,0,k_4}$ have the structure of infinite dimensional Hilbert spaces \cite{IJMPA} with inner product
\[ (u,v)=\int_{p\in({\bf R}^4)^{k_4}}\delta_{\alpha\beta}(u^{\alpha}(p))^{*}v^{\beta}(p)\,dp, \]
where $\delta=\bigotimes_{r=0}^{k_2}\delta$ and $dp=dp_1\cdots dp_{k_4}$.              

The space ${\mathcal P}_{k_1,k_2,k_3,k_4}$ is a subspace of the dual space ${\mathcal F}^{*}_{k_1,k_2,k_3,k_4}$.

\section{Scattering and other interaction processes}

\subsection{Covariant operators}

We consider scattering and other interaction processes to be associated with linear maps from a space of multiparticle states at a point $x\in X$ to a dual space of multiparticle states at the point $x$. Consider the electroweak sector. It can be shown (c.f. \cite{IJMPA}, Section 3.4, p. 9) that such maps in this case may be induced by operators ${\mathcal M}:{\mathcal F}_{k_1,k_2,k_3,k_4}\rightarrow{\mathcal P}_{k_1^{\prime},k_2^{\prime},k_3^{\prime}k_4^{\prime}}$ which intertwine with the action of $U(2)$ on the $U(2)$ bundle constructed in Section~\ref{section:gauge_groups}. We call such operators ($U(2)$) covariant. Covariant operators are invariants of the geometry of space-time \cite{IJMPA}.

A particular class of operators is generated by kernels ${\mathcal M}:\{0,1,2,3\}^{k_1^{\prime}}\times\{1,2\}^{k_2^{\prime}}\times\{0,1,2,3\}^{k_3^{\prime}}\times\{0,1,2,3\}^{k_1}\times\{1,2\}^{k_2}\times\{0,1,2,3\}^{k_3}\times({\bf R}^4)^{k_4^{\prime}}\times({\bf R}^4)^{k_4}\rightarrow{\bf C}$ which are smooth and polynomially bounded in their continuous arguments.
The kernels of this type act on Fock space according to
\[ ({\mathcal M}u)^{\mu^{\prime}\alpha^{\prime}i^{\prime}}(p^{\prime})=\int_{p\in({\bf R}^4)^{k_4}}{\mathcal M}^{\mu^{\prime}\alpha^{\prime}i^{\prime}}{}_{\mu\alpha i}(p^{\prime},p)u^{\mu\alpha i}(p)\,dp, \]
where $dp=dp_1\cdots dp_{k_4}$.

A more general class of kernels consists of kernele ${\mathcal M}:\{0,1,2,3\}^{k_1^{\prime}}\times\{1,2\}^{k_2^{\prime}}\times\{0,1,2,3\}^{k_3^{\prime}}\times\{0,1,2,3\}^{k_1}\times\{1,2\}^{k_2}\times\{0,1,2,3\}^{k_3}\times({\bf R}^4)^{k_4^{\prime}}\times({\mathcal B}_0({\bf R}^4))^{k_4}\rightarrow{\bf C}$ which are smooth and polynomially bounded in their continuous arguments and Borel measures in their set arguments (here ${\mathcal B}_0({\bf R}^4)$ denotes the collection of relatively compact Borel subsets of Minkowski space ${\bf R}^4$). 
Kernels of this more general type act on Fock space according to
\[ ({\mathcal M}u)^{\mu^{\prime}\alpha^{\prime}i^{\prime}}(p^{\prime})=\int_{p\in({\bf R}^4)^{k_4}}{\mathcal M}^{\mu^{\prime}\alpha^{\prime}i^{\prime}}{}_{\mu\alpha i}(p^{\prime},dp)u^{\mu\alpha i}(p). \]

A mapping ${\mathcal M}:{\bf R}^4\rightarrow{\bf C}^{2\times2}$ will be said to be $U(2)$ covariant, or simply covariant, if
\[ {\mathcal M}(ap)=a{\mathcal M}(p)a^{-1},\forall a\in U(2),p\in{\bf R}^4. \]

There is, associated with any $\kappa=\hbox{diag}(a,a^{\dagger-1})\in K$ a Lorentz transformation $\Lambda(\kappa)$ given by $\Lambda(M(p))=aM(p)a^{\dagger}$ (see Refs. \cite{AMP,IJMPA}). In particular, when $a\in U(2)$ the map $(\Lambda(a))(p)=aM(p)a^{\dagger}=aM(p)a^{-1}$ is a rotation.
A matrix valued map ${\mathcal M}^{\mu}:{\bf R}^4\rightarrow{\bf C}^{2\times2},\mu\in\{0,1,2,3\}$ with one Lorentz index will be said to be $U(2)$ covariant, if
\[ {\mathcal M}^{\mu}(ap)=\Lambda(a)^{\mu}{}_{\nu}a{\mathcal M}^{\nu}(p)a^{-1},\forall a\in U(2),p\in{\bf R}^4, \]
where $\Lambda=\Lambda(a)$ is the rotation corresponding to $a\in U(2)$.

A kernel ${\mathcal M}:\{0,1,2,3\}^{k_1^{\prime}}\times\{1,2\}^{k_2^{\prime}}\times\{0,1,2,3\}^{k_1}\times\{1,2\}^{k_2}\times({\bf R}^4)^{k_4^{\prime}}\times({\bf R}^4)^{k_4}\rightarrow{\bf C}$ with no $u(2)$ indices is covariant (intertwining \cite{IJMPA}) if
\begin{align*}
{\mathcal M}^{\mu^{\prime}\alpha^{\prime}}{}_{\mu\alpha}(ap^{\prime},ap)=\Lambda^{\mu^{\prime}}{}_{\rho^{\prime}}\Lambda^{-1}{}_{\mu}{}^{\sigma}a^{\alpha^{\prime}}{}_{\beta^{\prime}}a^{-1}{}_{\alpha}{}^{\beta}{\mathcal M}^{\rho^{\prime}\beta^{\prime}}{}_{\sigma\beta}(p^{\prime},p), 
\end{align*}
for all $a\in U(2),p^{\prime}\in({\bf R}^4)^{k_4^{\prime}},p\in({\bf R}^4)^{k_4},\Lambda=\Lambda(a)$ and all free indices.

Given two kernels ${\mathcal M}_1^{\mu},{\mathcal M}_2^{\mu}:\{1,2\}^2\times({\bf R}^4)^2\rightarrow{\bf C}$, each with one Lorentz index but no $u(2)$ indices, we can define a kernel ${\mathcal M}=\eta_{\mu\nu}{\mathcal M}_1^{\mu}\otimes{\mathcal M}_2^{\nu}:\{1,2\}^4\times({\bf R}^4)^4\rightarrow{\bf C}$ by
\begin{align*} 
{\mathcal M}^{\alpha_1^{\prime}\alpha_2^{\prime}}{}_{\alpha_1\alpha_2}(p_1^{\prime},p_2^{\prime},p_1,p_2)&=\eta_{\mu\nu}({\mathcal M}_1^{\mu}\otimes{\mathcal M}_2^{\nu})^{\alpha_1^{\prime}\alpha_2^{\prime}}{}_{\alpha_1\alpha_2}(p_1^{\prime},p_2^{\prime},p_1,p_2)\\
&=\eta_{\mu\nu}{\mathcal M}_1^{\mu\alpha_1^{\prime}}{}_{\alpha_1}(p_1^{\prime},p_1){\mathcal M}_2^{\nu\alpha_2^{\prime}}{}_{\alpha_2}(p_2^{\prime},p_2).
\end{align*}
If ${\mathcal M}_1,{\mathcal M}_2$ are covariant then ${\mathcal M}$ is covariant. 

If a kernel has $u(2)$ indices then $U(2)$ acts on it by means of the adjoint representation of $U(2)$ in $u(2)$.

\subsection{Dirac spinors}

Dirac spinors are complex 4-vector valued functions on momentum space associated with plane wave solutions of the Dirac equation. A plane wave ${\bf C}^4$ valued function on Minkowski space has the form
\[ \psi(x)=e^{-ip\cdot x}u, \]
where $p\in{\bf R}^4,u\in{\bf C}^4$. Such a wave function satisfies the Dirac equation if and only if
\begin{equation}\nonumber
(i{\slas\partial}-m)\psi=0\Leftrightarrow({\slas p}-m)u=0\Leftrightarrow{\slas p}u=mu. 
\end{equation}
It can be shown that the general form of a (positive energy) Dirac spinor is of the form
\begin{equation}\nonumber
u=\left(\begin{array}{c}
(p\cdot\sigma)^{\frac{1}{2}}\xi\\
(p\cdot{\overline\sigma})^{\frac{1}{2}}\xi\end{array}\right),
\end{equation}
where $\sigma=(\sigma_0,\sigma_1,\sigma_2,\sigma_3)^{\dagger},\overline{\sigma}=(\sigma_0,-\sigma_1,-\sigma_2,-\sigma_3)^{\dagger}$ and $\xi\in{\bf C}^2$.
Thus a basis for the space of Dirac spinors is given by $\{u(p,\alpha)\}_{\alpha=1,2,p^2=m^2}$ where
\[ u(p,\alpha)=\left(\begin{array}{c}
(p\cdot\sigma)^{\frac{1}{2}}e_{\alpha}\\
(p\cdot{\overline\sigma})^{\frac{1}{2}}e_{\alpha}\end{array}\right), \]
for $\alpha=1,2$ where $\{e_1,e_2\}$ is the standard basis for ${\bf C}^2$.

\subsection{$U(2)$ covariance and $K$ covariance}

$U(2)$ embeds in $K$ according to 
\[ a\mapsto\left(\begin{array}{cc}
a&0\\
0&a
\end{array}\right). \]
$K$ acts on ${\bf C}^{4\times4}$ \cite{IJMPA} and induces an action of $U(2)$ on ${\bf C}^{4\times4}$.
A map ${\mathcal M}:{\bf R}^4\rightarrow{\bf C}^{4\times4}$ is $U(2)$ covariant if
\[ {\mathcal M}(ap)=a{\mathcal M}(p)a^{-1},\forall a\in U(2),p\in{\bf R}^4. \]
We then, trivially, have the following
\newtheorem{lemma}{Lemma}
\begin{lemma} Let ${\mathcal M}:{\bf R}^4\rightarrow{\bf C}^{4\times4}$. Then if ${\mathcal M}$ is $K$ covariant then it is $U(2)$ covariant.
\end{lemma}
Thus, in this sense, $K$ covariance is a stronger condition that $U(2)$ covariance. The simple external vertex and the general external vertex described in the next two subsections are objects which manifest $U(2)$ covariance but not $K$ covariance,

\subsection{The simple external vertex\label{section:elementary_vertex}}
The above definitions apply also for kernels defined on subsets $S$ of Minkowski space which satisfy $O(1,3)^{\uparrow+}S=S$, or products of such sets.

Consider the Feynman amplitude fragment consisting of the kernel ${\mathcal M}^{\mu}:(H_{m^{\prime}}\times H_m)\rightarrow{\bf C}^{2\times2}$ with one Lorentz index, given by
\[ {\mathcal M}^{\mu}{}^{\alpha^{\prime}}{}_{\alpha}(p^{\prime},p)=\overline{u}(p^{\prime},\alpha^{\prime})\gamma^{\mu}u(p,\alpha), \]
where 
\[ H_m=H_m^{+}=\{p\in{\bf R}^4:p^2=m^2,p^0>0\},H_m^{-}=\{p\in{\bf R}^4:p^2=m^2,p^0<0\},m\geq0, \]
are the positive energy and negative energy mass $m$ mass shells. Then we have the following.
\newtheorem{theorem}{Theorem}
\begin{theorem}\label{th:elem_vertex_thm}
${\mathcal M}^{\mu}$ is a $U(2)$ covariant kernel with one Lorentz index.
\end{theorem}
{\bf Proof}
\begin{align*}
{\mathcal M}^{\mu\alpha^{\prime}}{}_{\alpha}(p^{\prime},p)&=\overline{u}(p^{\prime},\alpha^{\prime})\gamma^{\mu}u(p,\alpha)\\
&=\overline{\left(\begin{array}{c}
(p^{\prime}\cdot\sigma)^{\frac{1}{2}}e_{\alpha^{\prime}}\\
(p^{\prime}\cdot\overline{\sigma})^{\frac{1}{2}}e_{\alpha^{\prime}}
\end{array}\right)}\left(\begin{array}{cc}
0&\sigma^{\mu}\\
\overline{\sigma}^{\mu}&0
\end{array}\right)
\left(
\begin{array}{c}
(p\cdot\sigma)^{\frac{1}{2}}e_{\alpha}\\
(p\cdot\overline{\sigma})^{\frac{1}{2}}e_{\alpha}
\end{array}
\right)\\
&=(e_{\alpha^{\prime}}^{\dagger}(p^{\prime}\cdot\sigma)^{\frac{1}{2}\dagger},e_{\alpha^{\prime}}^{\dagger}(p^{\prime}\cdot\overline{\sigma})^{\frac{1}{2}\dagger})\left(\begin{array}{cc}
0&1\\
1&0
\end{array}\right)\left(\begin{array}{cc}
0&\sigma^{\mu}\\
\overline{\sigma}^{\mu}&0
\end{array}\right)\left(
\begin{array}{c}
(p\cdot\sigma)^{\frac{1}{2}}e_{\alpha}\\
(p\cdot\overline{\sigma})^{\frac{1}{2}}e_{\alpha}
\end{array}\right).
\end{align*}
It is straightforward to verify that the eigenvalues of the hermitian matrix $p^{\prime}\cdot\sigma$ are all positive for all $p^{\prime}\in H_m$ (see Appendix A). Therefore the  matrix $p^{\prime}\cdot\sigma$ has a unique positive definite hermitian square root for all $p^{\prime}\in H_m$.  Hence
\begin{align*}
{\mathcal M}^{\mu\alpha^{\prime}}{}_{\alpha}(p^{\prime},p)&=(e_{\alpha^{\prime}}^{\dagger}(p^{\prime}\cdot\sigma)^{\frac{1}{2}},e_{\alpha^{\prime}}^{\dagger}(p^{\prime}.\overline{\sigma})^{\frac{1}{2}})\left(\begin{array}{cc}
\overline{\sigma}^{\mu}&0\\
0&\sigma^{\mu}
\end{array}\right)\left(
\begin{array}{c}
(p\cdot\sigma)^{\frac{1}{2}}e_{\alpha}\\
(p\cdot\overline{\sigma})^{\frac{1}{2}}e_{\alpha}
\end{array}\right)\\
&=e_{\alpha^{\prime}}^{\dagger}(p^{\prime}\cdot\sigma)^{\frac{1}{2}}\overline{\sigma}^{\mu}(p\cdot\sigma)^{\frac{1}{2}}e_{\alpha}+e_{\alpha^{\prime}}^{\dagger}(p^{\prime}\cdot\overline{\sigma})^{\frac{1}{2}}\sigma^{\mu}(p\cdot\overline{\sigma})^{\frac{1}{2}}e_{\alpha}\\
&=[(p^{\prime}\cdot\sigma)^{\frac{1}{2}}\overline{\sigma}^{\mu}(p\cdot\sigma)^{\frac{1}{2}}+(p^{\prime}\cdot\overline{\sigma})^{\frac{1}{2}}\sigma^{\mu}(p\cdot\overline{\sigma})^{\frac{1}{2}}]^{\alpha^{\prime}}{}_{\alpha}.
\end{align*} 
Thus we can write
\[ {\mathcal M}^{\mu}(p^{\prime},p)=(p^{\prime}\cdot\sigma)^{\frac{1}{2}}\overline{\sigma}^{\mu}(p\cdot\sigma)^{\frac{1}{2}}+(p^{\prime}\cdot\overline{\sigma})^{\frac{1}{2}}\sigma^{\mu}(p\cdot\overline{\sigma})^{\frac{1}{2}}. \]
Now we have the following
\begin{lemma}\label{lemma:covariance_2}
Let $f:{\bf R}^4\rightarrow{\bf C}^{2\times2}$ be defined by
\begin{equation}\label{eq:R4_u2}
f(p)=p\cdot\sigma.
\end{equation}
Then $f$ is $U(2)$ covariant.
\end{lemma}
{\bf Proof}
In general if $f:{\bf R}^4\rightarrow{\bf C}^{2\times2}$ we have that $f$ is $U(2)$ covariant if and only if
\[ f(ap)=af(p)a^{-1}, \forall p\in {\bf R}^4,a \in U(2), \]
Now we know from \cite{AMP,IJMPA} that
\[ \Sigma(\kappa p)=\kappa\Sigma(p)\kappa^{-1},\forall\kappa\in K,p\in{\bf R}^4, \]
where $\Sigma:{\bf R}^4\rightarrow su(2,2)$ is the Feynman slash $\Sigma(p)={\slas p}$. This is true, in particular, when
\[ \kappa=\left(\begin{array}{cc}
a&0\\
0&a\end{array}\right),a\in U(2). \]
Therefore, since, through the standard identification of ${\bf R}^4$ with $iu(2)$, we have $ap=aM(p)a^{\dagger}=\kappa p$, and so
\begin{align*}
\left(\begin{array}{cc}
0&(ap)\cdot\sigma\\
(ap)\cdot\overline{\sigma}&0
\end{array}\right)&=\left(\begin{array}{cc}
0&(\kappa p)\cdot\sigma\\
(\kappa p)\cdot\overline{\sigma}&0
\end{array}\right)=\left(\begin{array}{cc}
a&0\\
0&a\end{array}\right)\left(\begin{array}{cc}
0&p\cdot\sigma\\
p\cdot\overline{\sigma}&0\end{array}\right)\left(\begin{array}{cc}
a^{-1}&0\\
0&a^{-1}\end{array}\right)\\
&=\left(\begin{array}{cc}
0&a(p\cdot\sigma)a^{-1}\\
a(p\cdot\overline{\sigma})a^{-1}&0
\end{array}\right)
\end{align*}
Therefore
$f(ap)=af(p)a^{-1},\forall a\in U(2),p\in{\bf R}^4$.
 $\Box$\\
Also we have shown in this proof that $p\mapsto p\cdot\overline{\sigma}$ is $U(2)$ covariant. We now require the following.
\begin{lemma}\label{lemma:square_root_1}
Let $A\in{\bf C}^{2\times2}$ be a positive definite hermitian matrix and $B\in{\bf C}^{2\times2}$ be a square root of $A$. Then $B$ is hermitian with eigenvalues $\xi_1\in\{\pm\lambda_1^{\frac{1}{2}}\},\xi_2\in\{\pm\lambda_2^{\frac{1}{2}}\}$ where $\lambda_1,\lambda_2$ are the eigenvalues of $A$.
\end{lemma}
{\bf Proof} By well known theorems $A$ can be diagonalized by a unitary transformation and $B$ can be placed into upper triangular form by a unitary transformation. Therefore there exists $U_1,U_2\in U(2),\xi_1,\xi_2,\xi_3\in{\bf C}$ such that
\[ A=U_1\left(\begin{array}{cc}
\lambda_1&0\\
0&\lambda_2
\end{array}\right)U_1^{-1}, \]
\[ B=U_2\left(\begin{array}{cc}
\xi_1&\xi_2\\
0&\xi_3
\end{array}\right)U_2^{-1}. \]
Thus
\[ A=B^2=U_2\left(\begin{array}{cc}
\xi_1&\xi_2\\
0&\xi_3
\end{array}\right)\left(\begin{array}{cc}
\xi_1&\xi_2\\
0&\xi_3
\end{array}\right)U_2^{-1}=U_2\left(\begin{array}{cc}
\xi_1^2&\xi_2(\xi_1+\xi_3)\\
0&\xi_3^2
\end{array}\right)U_2^{-1}, \]
and thus
\[ \left(\begin{array}{cc}
\lambda_1&0\\
0&\lambda_2
\end{array}\right)=U_1^{-1}AU_1=U_1^{-1}U_2\left(\begin{array}{cc}
\xi_1^2&\xi_2(\xi_1+\xi_3)\\
0&\xi_3^2
\end{array}\right)U_2^{-1}U_1. \]
Hence $\xi_2=0$ or $\xi_3=-\xi_1$. If $\xi_3=-\xi_1$ then $A$ is a scalar matrix and the required result is trivial. So we can assume that $\xi_2=0$ in which case $B$ is seen to be hermitian such that its eigenvalues squared are the eigenvalues of $A$. $\Box$
\begin{lemma}
Let $f(p)=(p\cdot\sigma)^{\frac{1}{2}},p\in H_m$. Then $f$ is $U(2)$ covariant.
\end{lemma}
{\bf Proof}
$f(p)^2=p\cdot\sigma$. Therefore, for all $a\in U(2),p\in H_m$, using Lemma~\ref{lemma:covariance_2},
\[ f(ap)^2=(ap)\cdot\sigma=a(p\cdot\sigma)a^{-1}=a(p\cdot\sigma)^{\frac{1}{2}}a^{-1}a(p\cdot\sigma)^{\frac{1}{2}}a^{-1}=(af(p)a^{-1})^2. \]
$p\cdot\sigma$ is hermitian with positive eigenvalues. Let these eigenvalues be $\lambda_1,\lambda_2>0$. Then there exists $U\in U(2)$ such that $p\cdot\sigma=U\mbox{diag}(\lambda_1,\lambda_2)U^{-1}$. Therefore $U\mbox{diag}(\pm\lambda_1^{\frac{1}{2}},\pm\lambda_2^{\frac{1}{2}})U^{-1}$ are square roots of $p\cdot\sigma$. Conversely, Let $B\in{\bf C}^{2\times2}$ be any square root of $(p\cdot\sigma)$. Then, by Lemma \ref{lemma:square_root_1}, $B$ is hermitian with eigenvalues $\xi_1\in\{\pm\lambda_1^{\frac{1}{2}}\},\xi_2\in\{\pm\lambda_2^{\frac{1}{2}}\}$ and so $B$ has the form $U\mbox{diag}(\xi_1,\xi_2)U^{-1}$ for $U=(v_1,v_2)\in U(2)$ in which $v_1,v_2$ are the eigenvectors of $p\cdot\sigma$ corresponding to the eigenvalues $\lambda_1,\lambda_2$.
Therefore both $f(ap)$ and $af(p)a^{-1}$ are of the form $U\mbox{diag}(\pm\lambda_1^{\frac{1}{2}},\pm\lambda_2^{\frac{1}{2}})U^{-1}$ for $U\in U(2)$ where $\lambda_1,\lambda_2$ are the eigenvalues of $(ap)\cdot\sigma$. Hence, since they both agree when $a=e=1_2$ and $U(2)$ is connected it follows, by continuity, that they agree for all $a\in U(2)$. Therefore
\[ f(ap)=af(p)a^{-1},\forall p\in H_m,a\in U(2). \]
$\Box$
\begin{lemma} \[ \Lambda^{\mu}{}_{\nu}a\sigma^{\nu}a^{-1}=\sigma^{\mu}\mbox{ and }\Lambda^{\mu}{}_{\nu}a\overline{\sigma}^{\nu}a^{-1}={\overline\sigma}^{\mu}\mbox{ for all }a\in U(2). \]
\end{lemma}
{\bf Proof}
We have the result that \cite{IJMPA}
\[ \kappa\gamma_{\mu}\kappa^{-1}=\Lambda^{\nu}{}_{\mu}\gamma_{\nu},\forall\kappa\in K, \]
where $\Lambda=\Lambda(\kappa)$ is the Lorentz transformation corresponding to $\kappa$. 
Thus
\begin{align*}
\Lambda^{\mu}{}_{\nu}\kappa\gamma^{\nu}\kappa^{-1}&=\eta^{\nu\alpha}\Lambda^{\mu}{}_{\nu}\kappa\gamma_{\alpha}\kappa^{-1}=\eta^{\nu\alpha}\Lambda^{\mu}{}_{\nu}\Lambda^{\beta}{}_{\alpha}\gamma_{\beta}=\eta^{\mu\beta}\gamma_{\beta}=\gamma^{\mu}.
\end{align*}
Therefore using the fact that
\[ \gamma^{\mu}=\left(\begin{array}{cc}
0&\sigma^{\mu}\\
\overline{\sigma}^{\mu}&0
\end{array}\right), \]
and taking 
\[ \kappa=\left(\begin{array}{cc}
a&0\\
0&a
\end{array}\right), \]
the required result follows. $\Box$

From these lemmas it is straightforward to show that Theorem~\ref{th:elem_vertex_thm} follows. $\Box$

\subsection{The general external vertex\label{section:general_vertex}}

The general external vertex is the map ${\mathcal M}^{\mu}:(H_{m^{\prime}}\times H_m)\rightarrow{\bf C}^{2\times2}$ given by
\begin{equation}\label{eq:general_vertex}
{\mathcal M}^{\mu}{}^{\alpha^{\prime}}{}_{\alpha}(p^{\prime},p)=\overline{u}(p^{\prime},\alpha^{\prime})\Theta^{\mu}(p^{\prime},p)u(p,\alpha),
\end{equation}
where $\Theta^{\mu}:(H_{m^{\prime}}\times H_m)\rightarrow{\bf C}^{4\times4}$. This is the general form of a QFT vertex with two external fermionic lines, from the point of view of particle momenta and polarizations.
\begin{theorem}
Let ${\mathcal M}$ be given by Eq.~\ref{eq:general_vertex}. Then if $\Theta$ is $U(2)$ covariant then ${\mathcal M}$ is covariant.
\end{theorem}
{\bf Proof}
Let
\[ \Theta^{\mu}=\left(\begin{array}{cc}
\Theta_1^{\mu}&\Theta_2^{\mu}\\
\Theta_3^{\mu}&\Theta_4^{\mu}
\end{array}\right). \]
Since $\Theta^{\mu}$ is $U(2)$ covariant we have
\[ \left(\begin{array}{cc}
\Theta_1^{\mu}(\Lambda p^{\prime},\Lambda p)&\Theta_2^{\mu}(\Lambda p^{\prime},\Lambda p)\\
\Theta_3^{\mu}(\Lambda p^{\prime},\Lambda p)&\Theta_4^{\mu}(\Lambda p^{\prime},\Lambda p)
\end{array}\right)=\Lambda^{\mu}{}_{\nu}\left(\begin{array}{cc}
a&0\\
0&a
\end{array}\right)\left(\begin{array}{cc}
\Theta_1^{\nu}(p^{\prime},p)&\Theta_2^{\nu}(p^{\prime},p)\\
\Theta_3^{\nu}(p^{\prime},p)&\Theta_4^{\nu}(p^{\prime},p)
\end{array}\right)\left(\begin{array}{cc}
a^{-1}&0\\
0&a^{-1}
\end{array}\right). \]
Thus
\begin{equation}\label{eq:general_U_2_covariance_property}
\Theta_i^{\mu}(\Lambda p^{\prime},\Lambda p)=\Lambda^{\mu}{}_{\nu}a\Theta_i^{\nu}(p^{\prime},p)a^{-1}, \forall\mu=0,1,2,3, a\in U(2),p^{\prime}\in H_{m^{\prime}},p\in H_m,i=1,\ldots4,
\end{equation}
where $\Lambda=\Lambda(a)$ is the rotation corresponding to $a$. Then
\begin{align*}
{\mathcal M}^{\mu\alpha^{\prime}}{}_{\alpha}(p^{\prime},p)&=\overline{u}(p^{\prime},\alpha^{\prime})\Theta^{\mu}u(p,\alpha)\\
&=\overline{\left(\begin{array}{c}
(p^{\prime}\cdot\sigma)^{\frac{1}{2}}e_{\alpha^{\prime}}\\
(p^{\prime}\cdot\overline{\sigma})^{\frac{1}{2}}e_{\alpha^{\prime}}
\end{array}\right)}\left(\begin{array}{cc}
\Theta_1^{\mu}&\Theta_2^{\mu}\\
\Theta_3^{\mu}&\Theta_4^{\mu}
\end{array}\right)
\left(
\begin{array}{c}
(p\cdot\sigma)^{\frac{1}{2}}e_{\alpha}\\
(p\cdot\overline{\sigma})^{\frac{1}{2}}e_{\alpha}
\end{array}
\right)\\
&=(e_{\alpha^{\prime}}^{\dagger}(p^{\prime}\cdot\sigma)^{\frac{1}{2}\dagger},e_{\alpha^{\prime}}^{\dagger}(p^{\prime}\cdot\overline{\sigma})^{\frac{1}{2}\dagger})\left(\begin{array}{cc}
0&1\\
1&0
\end{array}\right)\left(\begin{array}{cc}
\Theta_1^{\mu}&\Theta_2^{\mu}\\
\Theta_3^{\mu}&\Theta_4^{\mu}
\end{array}\right)\left(
\begin{array}{c}
(p\cdot\sigma)^{\frac{1}{2}}e_{\alpha}\\
(p\cdot\overline{\sigma})^{\frac{1}{2}}e_{\alpha}
\end{array}\right)\\
&=(e_{\alpha^{\prime}}^{\dagger}(p^{\prime}\cdot\sigma)^{\frac{1}{2}},e_{\alpha^{\prime}}^{\dagger}(p^{\prime}.\overline{\sigma})^{\frac{1}{2}})\left(\begin{array}{cc}
\Theta_3^{\mu}&\Theta_4^{\mu}\\
\Theta_1^{\mu}&\Theta_2^{\mu}
\end{array}\right)\left(
\begin{array}{c}
(p\cdot\sigma)^{\frac{1}{2}}e_{\alpha}\\
(p\cdot\overline{\sigma})^{\frac{1}{2}}e_{\alpha}
\end{array}\right)\\
&=e_{\alpha^{\prime}}^{\dagger}[((p^{\prime}\cdot\sigma)^{\frac{1}{2}}\Theta_3^{\mu}+(p^{\prime}\cdot\overline{\sigma})^{\frac{1}{2}}\Theta_1^{\mu},
(p^{\prime}\cdot\sigma)^{\frac{1}{2}}\Theta_4^{\mu}+(p^{\prime}\cdot\overline{\sigma})^{\frac{1}{2}}\Theta_2^{\mu})]
\left(\begin{array}{c}
(p\cdot\sigma)^{\frac{1}{2}}e_{\alpha}\\
(p\cdot\overline{\sigma})^{\frac{1}{2}}e_{\alpha}
\end{array}\right)\\
&=e_{\alpha^{\prime}}^{\dagger}[((p^{\prime}\cdot\sigma)^{\frac{1}{2}}\Theta_3^{\mu}+(p^{\prime}\cdot\overline{\sigma})^{\frac{1}{2}}\Theta_1^{\mu})(p\cdot\sigma)^{\frac{1}{2}}+
((p^{\prime}\cdot\sigma)^{\frac{1}{2}}\Theta_4^{\mu}+(p^{\prime}\cdot\overline{\sigma})^{\frac{1}{2}}\Theta_2^{\mu})(p\cdot{\overline\sigma})^{\frac{1}{2}}]e_{\alpha}\\
&=[((p^{\prime}\cdot\sigma)^{\frac{1}{2}}\Theta_3^{\mu}+(p^{\prime}.\overline{\sigma})^{\frac{1}{2}}\Theta_1^{\mu})(p\cdot\sigma)^{\frac{1}{2}}+
((p^{\prime}\cdot\sigma)^{\frac{1}{2}}\Theta_4^{\mu}+(p^{\prime}\cdot\overline{\sigma})^{\frac{1}{2}}\Theta_2^{\mu})(p\cdot{\overline\sigma})^{\frac{1}{2}}]^{\alpha^{\prime}}{}_{\alpha}
\end{align*}
Therefore we have
\[ {\mathcal M}^{\mu}(p^{\prime},p)=((p^{\prime}\cdot\sigma)^{\frac{1}{2}}\Theta_3^{\mu}+(p^{\prime}\cdot\overline{\sigma})^{\frac{1}{2}}\Theta_1^{\mu})(p\cdot\sigma)^{\frac{1}{2}}+
((p^{\prime}\cdot\sigma)^{\frac{1}{2}}\Theta_4^{\mu}+(p^{\prime}\cdot\overline{\sigma})^{\frac{1}{2}}\Theta_2^{\mu})(p\cdot{\overline\sigma})^{\frac{1}{2}}. \]
Therefore using Eq.~\ref{eq:general_U_2_covariance_property} and arguing as in the previous section we can show that ${\mathcal M}^{\mu}$ is $U(2)$ covariant. $\Box$

\section{The strong force sector, the complete Fock space and $SU(3)$ covariance}

We consider the Fock spaces ${\mathcal F}_{k_1,k_2,k_3,k_4,k_5,k_6}$ obtained by augmenting the electroweak Fock space tensor algebra by allowing our complex tensors to have additional indices, $a\in\{1,\ldots,8\}^{k_1}$ gluon color ($su(3)$) indices and $\rho\in\{1,2,3\}^{k_3}$ quark color ($SU(3)$) indices in addition to Lorentz indices $\mu\in\{0,1,2,3\}^{k_2}$, polarization indices $\alpha\in\{0,1\}^{k_4}$ and $u(2)$ indices $i\in\{0,1,2,3\}^{k_5}$. 

Thus a multiparticle state can be thought of as a complex Schwartz tensor valued function $p\mapsto u^{a\mu\rho\alpha i}(p)\in{\bf C}$ where $p\in({\bf R}^4)^{k_6}$. 

The transition functions $\Gamma_{2}$ and $U_{2}$ arise from the geometry of space-time as described in Section~\ref{section:gauge_groups}. $SU(3)$ acts on state vectors in the strong force sector.

One can define when a general kernel of the form ${\mathcal M}^{a^{\prime}\mu^{\prime}\rho^{\prime}\alpha^{\prime}i^{\prime}}{}_{a\mu\rho\alpha i}(p^{\prime},p)$ is covariant. However, in the present paper we will concentrate on the electroweak sector.

\section{Propagators as $K$ covariant measures}

Consider the electroweak boson propagators. Let $S\subset{\bf R}^4$ be some open set (or, more generally, a Borel set) such that $O(1,3)^{\uparrow+}S=S$. A tensor valued measure $\Pi^{\mu\nu}:{\mathcal B}_0(S)\rightarrow{\bf C}$  is Lorentz covariant if \cite{IJMPA,NPB}
\[ \Pi^{\mu\nu}(\Lambda\Gamma)=\Lambda^{\mu}{}_{\rho}\Lambda^{\nu}{}_{\sigma}\Pi^{\rho\sigma}(\Gamma),\forall\Lambda\in O(1,3)^{\uparrow+},\Gamma\in{\mathcal B}_0(S). \]

The on shell photon propagator can be viewed as the $K$ covariant Borel complex measure $D_{1,\mu\nu}^{\pm}:{\mathcal B}_0({\bf R}^4)\rightarrow{\bf C}$ given by
\[ D_{1,\mu\nu}^{\pm}=i\pi\eta_{\mu\nu}\Omega_0^{\pm}, \]
where $\Omega_m^{\pm}$ the standard Lorentz invariant measure concentrated on the mass shell $H_m^{\pm},m\geq0$ (see Ref.~\cite{IJMPA}).

The off shell photon propagator is the $K$ covariant Borel signed measure \cite{Halmos} $D_{2,\mu\nu}:{\mathcal B}_0(S)\rightarrow{\bf R}$ given by
\[ D_{2,\mu\nu}(\Gamma)=-\eta_{\mu\nu}\int_{\Gamma}\frac{dq}{q^2}, \]
where $S=\{q\in{\bf R}^4:q^2\neq0\})$.

The on shell part of the propagator for a vector boson of mass $M>0$ is the $K$ covariant Borel complex measure $D_{M,1,\mu\nu}^{\pm}:{\mathcal B}_0({\bf R}^4)\rightarrow{\bf C}$ given by
\[ D_{M,1,\mu\nu}^{\pm}(\Gamma)=i\pi\int_{\Gamma}(\eta_{\mu\nu}-M^{-2}q_{\mu}q_{\nu})\,\Omega_M^{\pm}(dq). \]
The off shell part is the covariant Borel signed measure $D_{M,2,\mu\nu}:{\mathcal B}_0(S)\rightarrow{\bf R}$ given by
\[ D_{M,2,\mu\nu}(\Gamma)=\int_{\Gamma}\frac{-\eta_{\mu\nu}+M^{-2}q_{\mu}q_{\nu}}{q^2-M^2}\,dq, \]
where $S=S_M=\{q\in{\bf R}^4:q^2\neq M^2\}$.

One can, similarly to the boson propagators, define the fermion propagators as $K$ covariant measures.

\section{The quantity $\Phi_{\mathcal M}=\overline{|{\mathcal M}|^2}$\label{section:Phi_M}}

\subsection{Lorentz invariance}

Given any Feynman amplitude ${\mathcal M}$ the computation of the quantity $|{\mathcal M}|^2$ results, because of the spinor outer product properties
\begin{align*}
\sum_{\alpha=1}^2u_{\alpha}(p)\overline{u}_{\alpha}(p)={\slas p}+m,\\
\sum_{\alpha=1}^2v_{\alpha}(p)\overline{v}_{\alpha}(p)={\slas p}-m,
\end{align*}
in a sum of products of $K$ covariant quantities. e.g. consider the case when factors are made up of terms of the form
\[ f(p_1,\ldots,p_n)=\mbox{tr}(({\slas p}_1\pm m)\cdots({\slas p}_n\pm m)), \]
Then 
\begin{align*}
f(\kappa p_1,\ldots,\kappa p_n)&=\mbox{tr}((\Sigma(\kappa p_1)\pm m)\cdots(\Sigma(\kappa p_n)\pm m))\\
&=\mbox{tr}(\kappa{\slas p}_1\kappa^{-1}\pm m)\cdots(\kappa{\slas p}_n\kappa^{-1}\pm m))\\
&=\mbox{tr}(\kappa({\slas p}_1\pm m)\kappa^{-1}\cdots\kappa({\slas p}_n\pm m)\kappa^{-1})\\
&=\mbox{tr}(\kappa({\slas p}_1\pm m)\cdots({\slas p}_n\pm m)\kappa^{-1})\\
&=\mbox{tr}(({\slas p}_1\pm m)\cdots({\slas p}_n\pm m))\\
&=f(p_1,\ldots,p_n),
\end{align*} 
for all $\kappa\in K$ where we have used the fundamental intertwining property of the Feynman slash \cite{AMP,IJMPA} which is that
\[ \Sigma(\kappa p)=\kappa\Sigma(p)\kappa^{-1}, \forall\kappa\in K,p\in{\bf R}^4, \]
in which $\Sigma(p)={\slas p}$. 

Inserted gamma matrices can be eliminated in a Lorentz invariant fashion using  gamma matrix commutation relations and contraction identities. Furthermore, propagators are all $K$ covariant. 

It is therefore fairly straightforward to give a general proof by induction of the Lorentz invariance of $\overline{|{\mathcal M}|^2}$.  Since the Lorentz invariance of $\overline{|{\mathcal M}|^2}$ is well known, we will not provide details of such a proof in this paper.

We will denote the Lorentz invariant object $\overline{|{\mathcal M}|^2}$ by $\Phi_{\mathcal M}$.

\subsection{Unitarity and probabilistic interpretation}

Consider the case of a $U(2)$ covariant kernel $p\mapsto{\mathcal M}^{\alpha^{\prime}}{}_{\alpha}(p)$ with one contravariant polarization index $\alpha^{\prime}\in\{1,2\}$, one covariant polarization index $\alpha\in\{1,2\}$ and one continuous argument $p\in H_m$ which is such that
\[ \int_{p\in H_m}\overline{|{\mathcal M}(p)|^2}\,\Omega_m(dp)<\infty. \]
The generalization to more general kernel types is straightforward.

${\mathcal M}$ is a matrix valued function on $H_m$. Let $X(p)={\mathcal M}^{\dagger}(p){\mathcal M}(p)$. $X$ is a matrix valued function on $H_m$. $X(p)$ is hermitian for all $p\in H_m$. Therefore there exists $U(p)\in U(2)$ such that $X(p)=U(p)\mbox{diag}(\lambda_1(p),\lambda_2(p))U(p)^{-1}$ where $\lambda_1(p),\lambda_2(p)\in{\bf R}$ are the eigenvalues of $X(p)$. By $U(2)$ covariance of ${\mathcal M}$ we have
\begin{equation}\label{eq:X_covariance}
X(ap)={\mathcal M}^{\dagger}(ap){\mathcal M}(ap)=(a{\mathcal M}(p)a^{-1})^{\dagger}a{\mathcal M}(p)a^{-1}=aX(p)a^{-1},\forall a\in U(2),p\in H_m.
\end{equation}
Thus
\[ X(U(p)^{-1}p)=U(p)^{-1}X(p)U(p)=\mbox{diag}(\lambda_1(p),\lambda_2(p)),\forall p\in H_m. \]
Therefore
\[ X(p)=\mbox{diag}(\lambda_1(U(p)p),\lambda_2(U(p)p)),\forall p\in H_m, \]
It is straightforward to show that the eigenvalues of a $U(2)$ covariant kernel are $U(2)$ invariant. Hence 
\[ X(p)=\mbox{diag}(\lambda_1(p),\lambda_2(p)),\forall p\in H_m, \] 
and so $X(p)$ is a diagonal matrix for all $p\in H_m$.

Let
\[ Y=\int_{H_m}X(p)\,\Omega_m(dp)\in{\bf C}^{4\times4}. \]
$Y$ is a diagonal hermitian matrix. 
We have, from Eq.~\ref{eq:X_covariance} that
\[ \int_{H_m}X(ap)\,\Omega_m(dp)=a\int_{H_m}X(p)\,\Omega_m(dp)\,a^{-1}, \]
But
\[ \int_{H_m}X(ap)\,\Omega_m(dp)=\int_{H_m}X(p)\,\Omega_m(dp). \]
Therefore
\begin{equation} \label{eq:commuting_condition}
Y=aYa^{-1},\forall a\in U(2).
\end{equation}
Clearly this condition will be satisfied for any $Y=\lambda,\lambda\in{\bf R}$. Conversely we will show that any diagonal hermitian matrix that satisfies Eq.~\ref{eq:commuting_condition} must be a real scalar matrix. Let $Y=\mbox{diag}(\lambda_1,\lambda_2)$ be a diagonal hermitian matrix that satisfies Eq.~\ref{eq:commuting_condition}. Taking $a=\left(\begin{array}{cc}
0&1\\
1&0
\end{array}\right)$ we deduce from the fact that $Ya=aY$ that $\lambda_2=\lambda_1$ and so $Y=\lambda$ for some $\lambda\in{\bf R}$.

Therefore the matrix $Y$ is, a real scalar matrix and the quantity
\begin{align*}
\frac{1}{2}\mbox{tr}(X(p)=\frac{1}{2}\mbox{tr}({\mathcal M}^{\dagger}(p){\mathcal M}(p))&=\frac{1}{2}\sum_{\alpha=1}^2({\mathcal M}^{\dagger}(p){\mathcal M}(p))_{\alpha\alpha}=\frac{1}{2}\sum_{\alpha,\beta=1}^2({\mathcal M}^{\dagger}(p))_{\alpha}{}^{\beta}{\mathcal M}(p)^{\beta}{}_{\alpha}\\
&=\frac{1}{2}\sum_{\alpha,\beta}({\mathcal M}(p)^{\beta}{}_{\alpha})^{*}{\mathcal M}(p)^{\beta}{}_{\alpha}=\overline{|{\mathcal M}(p)|^2}, 
\end{align*}
is non-negative and integrates to the value
\[ \int_{H_m}\frac{1}{2}\mbox{tr}(X(p))\,\Omega_m(dp)=\frac{1}{2}\mbox{tr}(\int_{H_m}X(p)\,\Omega_m(dp))=\frac{1}{2}\mbox{tr}(Y)=\lambda, \]
 Thus $\lambda\geq0$. The case $\lambda=0$ is trivial and corresponds to the case ${\mathcal M}=0$ so need not be considered. If we normalize $\Phi_{\mathcal M}=\overline{|{\mathcal M}|^2}$ by dividing by $\lambda$ then it integrates to 1 and has the interpretation of a probability density. Physical arguments can be given relating this quantity to the differential cross section for the process with Feynman amplitude ${\mathcal M}$ (e.g.Schwartz, 2018 \cite{Schwartz}, p. 59).

\section{Computation using spectral analysis of the masses of the elementary particles \label{section:fundamental_particle_masses}}

\subsection{Description of the method\label{section:description_of_method}}

For the rest of this paper we will concentrate on the case where $k_6=k_6^{\prime}=2$. This correspends to scattering of two particles or, more generally, to a process with two incoming particles and two outgoing particles. By crossing symmetry this includes decay of one particle into three particles.

Consider a QFT process described by a covariant kernel ${\mathcal M}$ which may involve charge, electromagnetic or color, or else be a weak interaction. In general ${\mathcal M}$ will depend on the masses $m_1,m_2,m_1^{\prime}$ and $m_2^{\prime}$ and the momenta $p_1,p_2,p_1^{\prime}$ and $p_2^{\prime}$ of the incoming and outgoing particles respectively. Consider the case when $m_1=m_2=m$ and $m_1^{\prime}=m_2^{\prime}=m^{\prime}$ (e.g. particle annihilation and production in the s channel). 

For concreteness assume that the incoming particles are a positron and an electron. Suppose that we have fixed the masses $m$ of the incoming positron and electron. The electron is a stable particle with infinite lifetime and a delta function peak in its mass spectrum. Fixing its mass amounts to taking the electron mass as the reference mass, which amounts to fixing the units in which mass is measured. This is equivalent to setting the electron mass $m_e=1$ to be unity (a convention which is frequently adopted in particle physics).

The quantity $\Phi=\Phi(p_1^{\prime},p_2^{\prime},p_1,p_2)=\Phi(m^{\prime},p_1^{\prime},p_2^{\prime},p_1,p_2)$ defined in the Section~\ref{section:Phi_M} depends implicitly on $m^{\prime}$ ($m$ is fixed and given). As described in that section we have, in general, that $\Phi=\Phi_{\mathcal M}$ is Lorentz invariant.
We have the following constraints which arise from conservation of energy-momentum and the on-shell property of incoming and outgoing particles:
\begin{equation}\label{eq:constraints}
p_1^{\prime}+p_2^{\prime}=p_1+p_2, p_1^2=p_2^2=m^2, p_1^{\prime2}=p_2^{\prime2}=m^{\prime2}. 
\end{equation}
It is a non-trivial problem in algebraic geometry to satisfy these constraints in an arbitrary frame. It is convenient, for the purpose of satisfying these constraints, and also to simplify the problem, to make a dynamic change to a CM frame using a Lorentz transformation $\Lambda=\Lambda(p_1,p_2)$ smoothly depending on the incoming momenta. Note that the CM frame is not unique but is defined up to multiplication by a rotation. 

In Appendix A we give an algorithm for computing a smooth dynamic transformation $(p_1,p_2)\mapsto\Lambda(p_1,p_2)$ to a CM frame for each $(p_1,p_2)\in(H_m)^2$.   

Let $\Lambda=\Lambda(p_1,p_2)$ and  $r_1=\Lambda p_1,r_2=\Lambda p_2$. Then, from the definition of what it means to be a CM frame, there exists $E=E(p_1,p_2)>0,{\vct p}={\vct p}(p_1,p_2)\in{\bf R}^3$ such that
\[ r_1=(E,{\vct p}),r_2=(E,-{\vct p}). \]
Let  $(r_1^{\prime},r_2^{\prime})\in({\bf R}^4)^2$ be any pair of output momenta in our CM frama. By conservation of momentum we must have
\[ r_1^{\prime}+r_2^{\prime}=r_1+r_2=(2E,{\vct0}). \]
Therefore there exists ${\vct p}^{\prime}\in{\bf R}^3,E^{\prime}>0$ such that
\[ r_1^{\prime}=(E^{\prime},{\vct p}^{\prime}),r_2^{\prime}=(E^{\prime},-{\vct p}^{\prime}). \]
But we must have $E^{\prime}+E^{\prime}=2E$ so $E^{\prime}=E$. Hence
\[ r_1^{\prime}=(E,{\vct p}^{\prime}),r_2^{\prime}=(E,-{\vct p}^{\prime}). \]

The differential cross section of a process with Feynman amplitude ${\mathcal M}$ is written in the literature on the subject as (c.f. (Schwartz \cite{Schwartz}, 2018, p. 61) (on rearranging)
\begin{align}\label{eq:diff_x_scn}
d\sigma&=\frac{1}{16}\frac{1}{(2\pi)^2}\frac{1}{|{\vct v}_2-{\vct v}_1|}\delta(\Sigma p)\frac{1}{E({\vct p}_1)}\frac{1}{E({\vct p}_2)}\frac{1}{E^{\prime}({\vct p}_1^{\prime})}\frac{1}{E^{\prime}({\vct p}_2^{\prime})}\overline{|{\mathcal M}|^2}\,d{\vct p}_1^{\prime}\,d{\vct p}_2^{\prime},  
\end{align}
where ${\vct v}_1$ and ${\vct v}_2$ are the velocities of the incoming particles (${\vct v}_1={\vct p}_1/p_1^0, {\vct v}_2={\vct p}_2/p_2^0$), $E({\vct p})=\omega_m({\vct p}),E^{\prime}({\vct p}^{\prime})=\omega_{m^{\prime}}({\vct p}^{\prime})$ (in which, for any ${\vct v}\in{\bf R}^3,m>0,\omega_m({\vct v})=(m^2+{\vct v}^2)^{\frac{1}{2}}$) and a $\delta$ function is included to enforce momentum conservation. 

It is somewhat problematic to interpret such a formula for two reasons. Firstly including the delta function makes $\frac{d\sigma}{d\Omega}$ into a distribution rather than a density function. Secondly, for any given set of values for the incoming momenta $p_1$ and $p_2$ there is a whole manifold $M(p_1,p_2)=\{(p_1^{\prime},p_2^{\prime}):p_1^{\prime}+p_2^{\prime}=p_1+p_2,p_1^{\prime2}=p_2^{\prime2}=m^{\prime2}\}$ of sets of values of the outgoing momenta $p_1^{\prime}$ and $p_2^{\prime}$ which satisy Eq.~\ref{eq:constraints}, that is the momentum conservation constraints together with the on-shell requirement constraints. There is, in general, for each $E_1^{\prime},E_2^{\prime}>0$, a non trivial submanifold $M_{(E_1^{\prime},E_2^{\prime})}(p_1,p_2)$ of $M(p_1,p_2)$ corresponding to energies $(E_1^{\prime},E_2^{\prime})$. The manifold $M_{(E_1^{\prime},E_2^{\prime})}(p_1,p_2)$ is not an open subset of $({\bf R}^4)^2$ thereby making it problematic to interpret the infinitesimal quantities $d{\vct p}_1^{\prime}$ and $d{\vct p}_2^{\prime}$ in Eq.~\ref{eq:diff_x_scn}.  

By carrying out the analysis in the dynamic CM frame $(p_1,p_2)\mapsto\Lambda(p_1,p_2)$ we can give an interpretation for Eq.~\ref{eq:diff_x_scn} and, moreover, determine a method for computing the masses of the elementary particles.

In usual applications $m^{\prime}$ is known (by experiment) and fixed but in our computations we leave $m^{\prime}$ as an unknown.

By Lorentz invariance of $\Phi$
\[ \Phi(m^{\prime},p_1^{\prime},p_2^{\prime},p_1,p_2)=\Phi(m^{\prime},r_1^{\prime},r_2^{\prime},r_1,r_2). \]

For given values of the incoming momenta $p_1$ and $p_2$ our CM frame is determined (Appendix A) and so $r_1$ and $r_2$ are determined. Thus $E=E(p_1,p_2)$ is determined. In fact 
\[ E(p_1,p_2)=r_1(p_1,p_2)^0=(\Lambda(p_1,p_2)p_1)^0. \]

For a given possible mass $m^{\prime}$ of the outgoing particles we must have $m^{\prime2}=E^2-{\vct p}^{\prime2}$. This has no solution if $E<m^{\prime}$. If $E\geq m^{\prime}$ then $|{\vct p}^{\prime}|=(E^2-m^{\prime2})^{\frac{1}{2}}$ and $r_1^{\prime}$ and $r_2^{\prime}$ must be of the form 
\begin{align*}
r_1^{\prime}&=(E,(E^2-m^{\prime2})^{\frac{1}{2}}\omega^{\prime})\\
r_2^{\prime}&=(E,-(E^2-m^{\prime2})^{\frac{1}{2}}\omega^{\prime}),
\end{align*}
for some $\omega^{\prime}\in S^2$ where $S^2\subset{\bf R}^3$ is the (surface of) the unit sphere.

We define the differential cross section (corresponding to Eq.~\ref{eq:diff_x_scn}) to be given by
\begin{align}
\left(\frac{d\sigma}{d\Omega}\right)_{\mbox{CM}}(m^{\prime},r_1,r_2,\omega^{\prime})&=\frac{1}{32}\frac{1}{(2\pi)^2}\frac{1}{|{\vct v}_2-{\vct v}_1|}\frac{1}{E(r_1)^2}\Phi_{\mathcal M}(m^{\prime},(E(r_1),(E(r_1)^2-m^{\prime2})^{\frac{1}{2}}\omega^{\prime}),\nonumber\\
&(E(r_1),-(E(r_1)^2-m^{\prime2})^{\frac{1}{2}}\omega^{\prime}),r_1,r_2)\nonumber\\
&=\frac{1}{|{\vct v}_2-{\vct v}_1|}\Xi(m^{\prime},r_1,r_2,\omega^{\prime})\label{eq:my_diff_x_scn},
\end{align}
where
\begin{align*}
\Xi(r_1,r_2,\omega^{\prime})=&\frac{1}{32}\frac{1}{(2\pi)^2}\frac{1}{E(r_1)^2}\Phi_{\mathcal M}(m^{\prime},(E(r_1),(E(r_1)^2-m^{\prime2})^{\frac{1}{2}}\omega^{\prime}),\\
&(E(r_1),-(E(r_1)^2-m^{\prime2})^{\frac{1}{2}}\omega^{\prime}),r_1,r_2),
\end{align*}
for $E(r_1)=r_1^0\geq m^{\prime}$. 

Thus, when written in this form, $\left(\frac{d\sigma}{d\Omega}\right)_{\mbox{CM}}$ is, for fixed $m^{\prime}$, a density on the set
\[ S=\{(r_1,r_2,\omega^{\prime})\in H_m\times H_m\times S^2:r_1^0,r_2^{\prime0}\geq m^{\prime2}\}. \]

In the CM frame, 
\begin{align*}
\frac{1}{|{\vct v}_2-{\vct v}_1|}&=\frac{1}{|{\vct p}_2/p_2^0-{\vct p}_1/p_1^0|}=\frac{p_1^0p_2^0}{|p_1^0{\vct p}_2-p_2^0{\vct p}_1|}=\frac{E^2}{|E(-{\vct p})-E{\vct p}|}=\frac{E}{2|{\vct p}|}=\frac{E}{2(E^2-m^2)^{\frac{1}{2}}},
\end{align*}
and so, in the high energy limit,
\[ \frac{1}{|{\vct v}_2-{\vct v}_1|}=\frac{1}{2}\frac{|{\vct p}_2|}{|{\vct p}_1|}=\frac{1}{2}, \]
in which case the differential cross section can be written as
\begin{align*}
\left(\frac{d\sigma}{d\Omega}\right)_{\mbox{CM,HE}}(m^{\prime},r_1,r_2,\omega^{\prime})&=\frac{1}{64}\frac{1}{(2\pi)^2}\frac{|{\vct p}_2|}{|{\vct p}_1|}\frac{1}{E(r_1)^2}\Phi_{{\mathcal M},\mbox{HE}}(m^{\prime},(E(r_1),E(r_1)\omega^{\prime}),\\
&(E(r_1),-E(r_1)\omega^{\prime}),r_1,r_2).
\end{align*}

In the CM coordinate system, since $r_1=(E,{\vct p})$ and $r_2=(E,-{\vct p})$ we can write, for given $m^{\prime}>0$,
\[ \left(\frac{d\sigma}{d\Omega}\right)_{\mbox{CM}}=\frac{1}{|{\vct v}_2-{\vct v}_1|}\Xi(E,{\vct p},\omega^{\prime}). \]
Furthermore, using ${\vct p}=(E^2-m^2)^{\frac{1}{2}}\omega$ we have
\[ \left(\frac{d\sigma}{d\Omega}\right)_{\mbox{CM}}=\frac{1}{|{\vct v}_2-{\vct v}_1|}\Xi(E,\omega,\omega^{\prime}). \]
The total (CM) cross section can be written as
\begin{equation}
\sigma=\sigma(E)=\frac{1}{|{\vct v}_2-{\vct v}_1|}\frac{1}{4\pi}\int_{\omega,\omega^{\prime}\in S^2}\Xi(E,\omega,\omega^{\prime})\,d\omega\,d\omega^{\prime},\label{eq:my_total_x_scn}
\end{equation}
where $d\omega$ denotes the area measure on $S^2$.  

The differential cross section and the total cross section are not Lorentz invariant quantities. It is preferable to focus on Lorentz invariant quantities.
Let $\sigma_{{\mathcal M},\mbox{inv,tot}}(p_1,p_2,\Gamma)$ denote the quantity defined, in an arbitrary frame, by
\begin{align*}
\sigma_{{\mathcal M},\mbox{inv,tot}}(p_1,p_2,\Gamma)=&\int_{|{\vct q}|\leq E(p_1,p_2)}\chi_{\Gamma}((E(p_1,p_2)^2-{\vct q}^2)^{\frac{1}{2}})\Phi_{\mathcal M}(m^{\prime},(E(p_1,p_2),{\vct q}),\\
&(E(p_1,p_2),-{\vct q}),r_1(p_1,p_2),r_2(p_1,p_2))\,d{\vct q}.
\end{align*}
In general $\sigma_{{\mathcal M},\mbox{inv,tot}}:H_m^2\times{\mathcal B}((0,\infty))\rightarrow[0,\infty]$ where ${\mathcal B}((0,\infty))$ is the Borel algebra of $(0,\infty)$.

Now $r_1$ and $r_2$ are on shell and $\pi(r_1)=-\pi(r_2)$ where $\pi:{\bf R}^4\rightarrow{\bf R}^3$ is the standard projection defined by $\pi(p)=\pi((p^0,{\vct p}))={\vct p}$. Also $r_1(p_1,p_2)^0=r_2(p_1,p_2)^0=E(p_1,p_2)$. Therefore
\begin{align*}
&r_1(p_1,p_2)=(E(p_1,p_2),{\vct r}(p_1,p_2)),\\
&r_2(p_1,p_2)=(E(p_1,p_2),-{\vct r}(p_1,p_2)),
\end{align*}
for some function $(p_1,p_2)\mapsto {\vct r}(p_1,p_2)\in{\bf R}^3$. 

It will be shown in a subsequent paper that $(p_1,p_2)\mapsto\sigma_{{\mathcal M},\text{inv,tot}}(p_1,p_2,\Gamma)$ is Lorentz invariant for all $\Gamma\in{\mathcal B}((0,\infty))$..

We now define the integrated total quantity $\sigma_{{\mathcal M}}:{\mathcal B}((0,\infty))\rightarrow[0,\infty]$ by
\begin{align}
\sigma_{{\mathcal M}}(\Gamma)=&\int\sigma_{{\mathcal M},\mbox{inv,tot}}(p_1,p_2,\Gamma)\,\Omega_{m}(dp_2)\,\Omega_{m}(dp_1),\label{eq:fundamental_fermion_formula}
\end{align}
$\sigma_{{\mathcal M}}:{\mathcal B}((0,\infty))\rightarrow[0,\infty]$ is simply the integral using Lorentz invariant integration of the Lorentz invariant function $\sigma_{{\mathcal M},\mbox{inv,tot}}$. ${\sigma}_{\mathcal M}$ is frame independent.

There is an issue with this definition when one integrates over all of $({\bf R}^4)^2$ (in a practical computer implementation one only integrates on a product of compact sets such as generalized rectangles). This issue, and another technical issue, is completely solved in our subsequent paper on particle mass computation which will follow this paper.

We call $\sigma_{{\mathcal M}}$ the integral mass spectrum of the outgoing particle family associated with ${\mathcal M}$. We propose that the peaks in the density function associated with $\sigma_{{\mathcal M}}$ are associated with the different elementary particles associated with ${\mathcal M}$. The location of the peaks on $(0,\infty)$ correspond to the masses of the elementary particles in the family.

\subsection{Calculation of the masses of charged leptons $\mu$ and $\tau$ \label{section:charged_leptons}}

\begin{figure} 
\centering
\includegraphics[width=6cm]{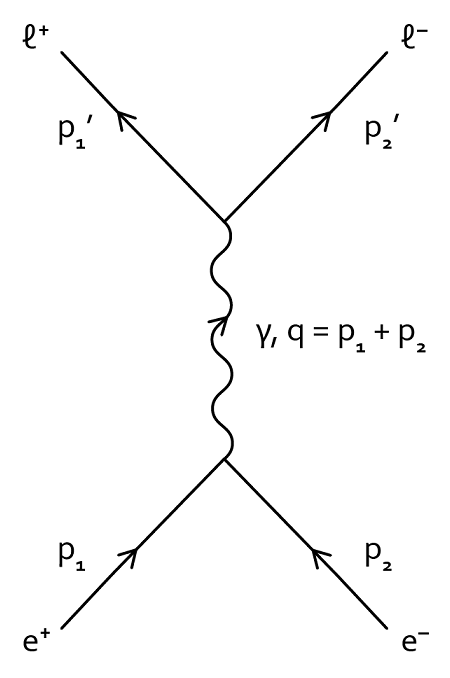}
\caption{Feynman diagram for the generation of charged leptons in QED\label{fig:charged_lepton_diagram}}
\end{figure}

\begin{figure} 
\centering
\includegraphics[width=15cm]{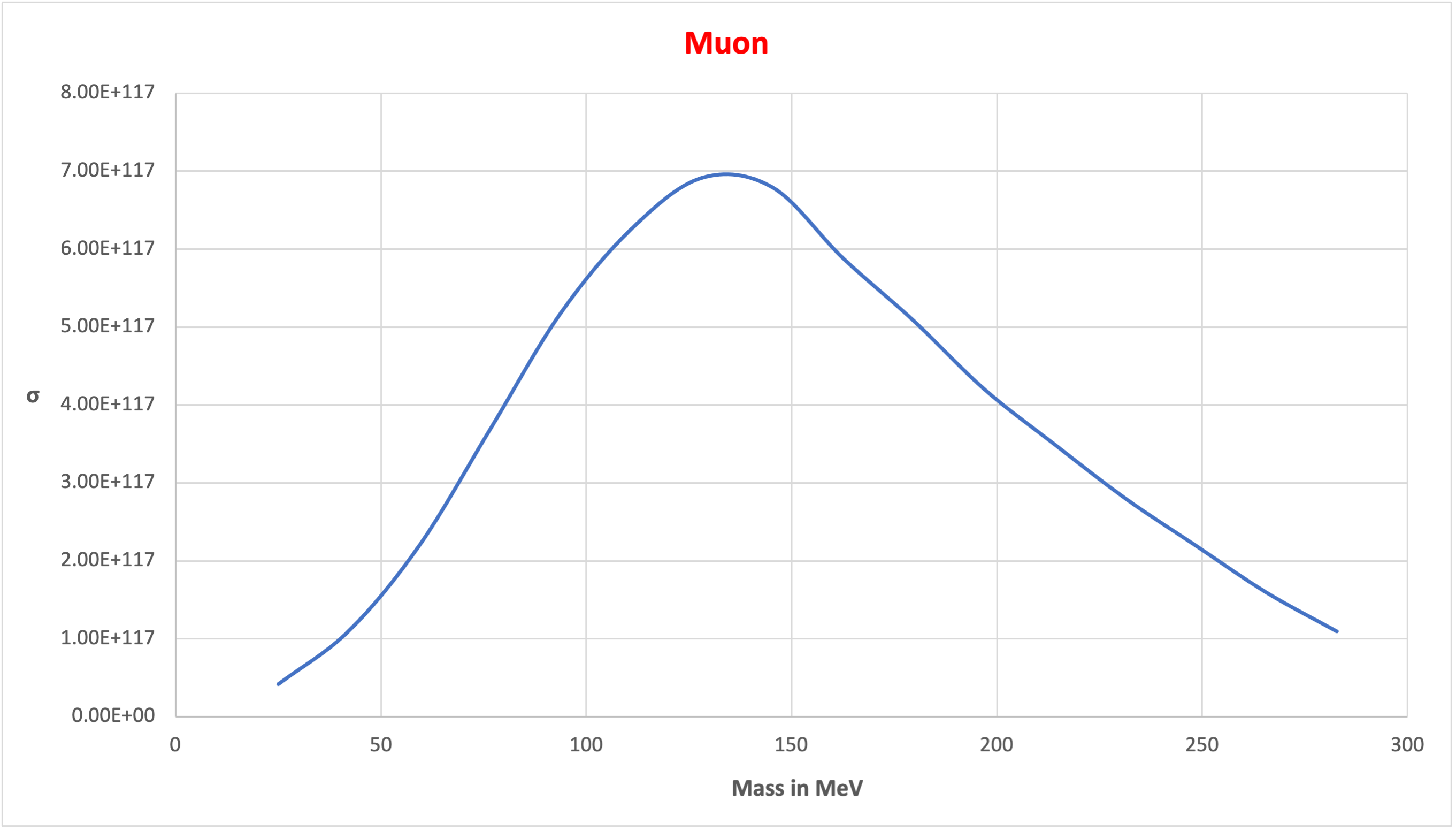}
\caption{Integral mass spectrum for charged lepton family in the viscinity of the muon mass \label{fig:muon}}
\end{figure}

Let ${\mathcal M}$ be the Feynman amplitude for the QED process $e^{+}e^{-}\rightarrow\gamma\rightarrow l^{+}l^{-}$ of charged lepton production associated with the Feynman diagram shown in Figure~\ref{fig:charged_lepton_diagram}. Using the Feynman rules ${\mathcal M}$ is given by
\begin{align*}
i{\mathcal M}=&\overline{v}(p_1,\alpha_1)i(-e)\gamma^{\mu}u(p_2,\alpha_2)iD_{\mu\nu}(p_1+p_2)\overline{u}(p_2^{\prime},\alpha_2^{\prime})i(-e)\gamma^{\nu}v(p_1^{\prime},\alpha_1^{\prime}).
\end{align*}
Assuming that $p_1$ and $p_2$ are on shell, $p_1+p_2$ is timelike. 
Therefore
\begin{equation}\nonumber
{\mathcal M}=e^2Q^{-2}\overline{v}(p_1,\alpha_1)\gamma^{\mu}u(p_2,\alpha_2)\eta_{\mu\nu}\overline{u}(p_2^{\prime},\alpha_2^{\prime})\gamma^{\nu}v(p_1^{\prime},\alpha_1^{\prime}),
\end{equation}
where
$Q=((p_1+p_2)^2)^{\frac{1}{2}}$.

One can readily compute that (Schwartz, 2018 \cite{Schwartz}, p. 232) 
\begin{equation}\nonumber
\overline{|{\mathcal M}|^2}=8e^4Q^{-4}((p_1\cdot p_1^{\prime})(p_2\cdot p_2^{\prime})+(p_2\cdot p_1^{\prime})(p_1\cdot p_2^{\prime})+m^2(p_1^{\prime}\cdot p_2^{\prime})+m^{\prime2}(p_1\cdot p_2)+2m^2m^{\prime2}),
\end{equation}
where $p_1,p_2\in H_m$ are the momenta of the incoming positron and electron respectively, $m$ is the mass of the electron, $p_1^{\prime},p_2^{\prime}\in H_{m^{\prime}}$ are the momenta of the outgoing antiparticle and particle respectively and $m^{\prime}$ is their mass. 

From this form for $\Phi_{\mathcal M}=\overline{|{\mathcal M}|^2}$ it is straightforward to compute from Eqns.~\ref{eq:my_diff_x_scn} and~\ref{eq:my_total_x_scn} that the differential cross section and total cross section in the extreme relativistic limit are, in the CM frame, given by the well known formulae
\begin{align*}
&\left(\frac{d\sigma}{d\Omega}\right)_{\mbox{CM}}=\frac{\alpha^2}{16E^2}(1+\cos^2(\theta)),\\
&\sigma=\frac{\pi\alpha^2}{3E^2}.
\end{align*}

Plugging the function $\Phi=\Phi_{\mathcal M}$ into the computations of the previous section and carrying out these integrals computationally we obtain an integral mass spectrum function $\sigma_{{\mathcal M}}$ whose values over the range from 0 MeV to 300 MeV is shown in Figure \ref{fig:muon}. 
It can be seen that there is a peak at $m^{\prime}\approx$ 100 MeV. The computation which was done on a 10 core machine with some degree of parallelization executed fairly rapidly. For this computation a number of input parameters to the program were used and these had the following values.
\begin{verbatim} 
double pi = 4.0 * atan(1.0);
double alpha = 1.0 / 137.0360; // fine structure constant
double e = sqrt(4.0 * pi * alpha); // electron charge in natural units
double m = 0.511; // electron mass in MeV
double Start = 0.0;  
double End = 300.0;
const double Lambda_integral = 200.0;
const int N_integral = 5;
const double delta_integral = Lambda_integral / N_integral;
const int N_int_angle = 5;
const int N_m_prime = 16;
const double delta_m_prime = (End - Start) / N_m_prime;
\end{verbatim}
From the data used to generate the figure (output from the program) the location of the peak is at $m=103.13$ MeV which differs by $2.4\%$ from the measured value of the mass $m=105.7$ MeV of the muon.

The C++ code for this calculation is given in Appendix B presented in the Supplementary Material for the paper as published in AIP Advances \cite{AIPAdv}. It is to be expected that with a higher value of N$\_$integral, integrating over a wider range and using high performance computing, that the peak location will coincide more closely with the measured value of $m_{\mu}=105.7$ MeV.

The only physical parameters input to this program are the electron mass, the fine structure constant and the electron charge. As described above we take the electron mass as the reference mass. The effect of the other two parameters is to multiply the integral mass spectrum by a constant. Therefore they are not relevant to determining the locations of the peaks in the integral mass spectrum, that is the particle masses

It should be noted that the exact result of this computation (and other computations described in this paper) depends on the values of the input parameters. If, for example, the parameter Lambda\_integral is varied significantly without increasing N\_integral then the peak shifts. This behavior is to be expected and can be remedied by increasing N\_integral.

\begin{figure} 
\centering
\includegraphics[width=15cm]{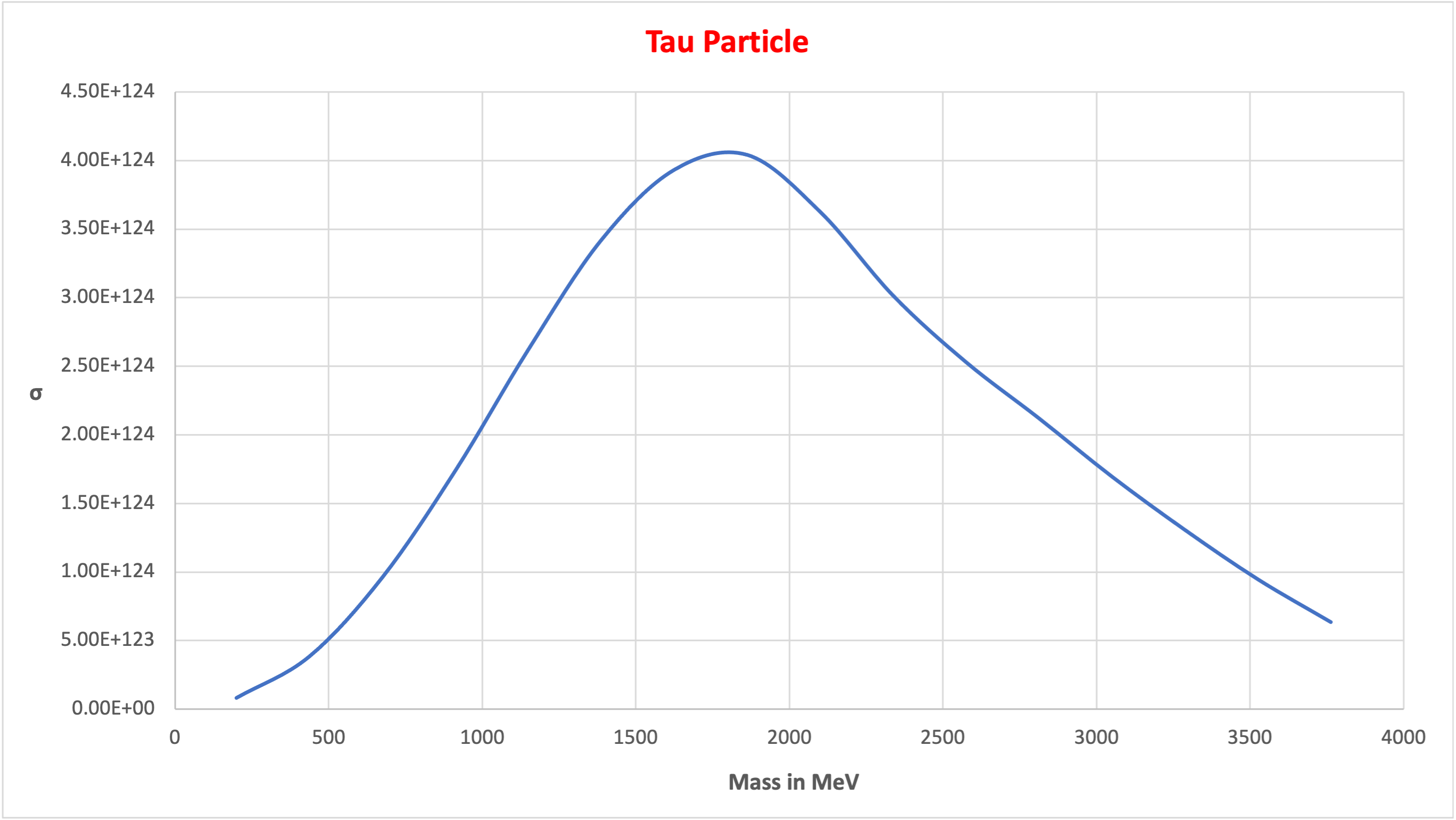}
\caption{Integral mass spectrum for the charged lepton family in the viscinity of the mass of the tau particle \label{fig:tau}}
\end{figure}

At higher energies there is another peak, that corresponding to the tau particle. The integral mass spectrum for the charged lepton family in the vicinity of the mass of the tau particle is shown in Figure~\ref{fig:tau}. The peak is located at a mass value of $m=1,746.9$ MeV which differs by $1.7\%$ from the measured value of $m_{\tau}$=1,777 MeV. The parameters input to the lepton mass computation program in this case were as follows.
\begin{verbatim}
double pi = 4.0 * atan(1.0);
double alpha = 1.0 / 137.0360; // fine structure constant
double e = sqrt(4.0 * pi * alpha); // electron charge in natural units
double m = 0.511; // electron mass in MeV
double Start = 200.0;  
double End = 3500.0; 
const double Lambda_integral = End;
const int N_integral = 6;
const double delta_integral = Lambda_integral / N_integral;
const int N_int_angle = 6;
const int N_m_prime = 16;
const double delta_m_prime = (End - Start) / N_m_prime;
\end{verbatim}

At still higher energies the integral mass spectrum of the charged lepton family seems to asymptote to zero (though it remains to prove this analytically). Thus we have established, computationally, that this family has three generations.

\subsection{Calculation of the mass of the Z$^0$ particle}

\begin{figure} 
\centering
\includegraphics[width=6cm]{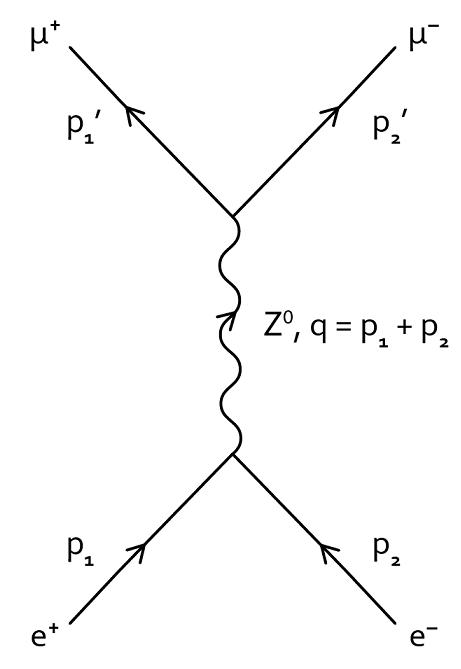}
\caption{Feynman diagram for the process $e^{+}e^{-}\rightarrow Z^0\rightarrow\mu^{+}\mu^{-}$ \label{fig:Z_boson_diagram}}
\end{figure}

Consider the process $e^{+}e^{-}\rightarrow Z^0\rightarrow \mu^{+}\mu^{-}$ of muon generation through the weak force whose Feynman diagram is shown in Figure~\ref{fig:Z_boson_diagram}. Using the Feynman rules the Feynman amplitude for this process is given by
\begin{align*}
i{\mathcal M}=&\overline{v}(p_1,\alpha_1)ig_W\gamma^{\mu}T^{\alpha}_{i_1i_2}u(p_2,\alpha_2)i\delta_{\alpha\alpha^{\prime}}D_{m_Z\mu\nu}(q)\overline{u}(p_2^{\prime},\alpha_2^{\prime})ig_W\gamma^{\nu}T^{\alpha^{\prime}}_{i_1^{\prime}i_2^{\prime}}v(p_1^{\prime},\alpha_1^{\prime}),
\end{align*}
where $g_W$ is the weak coupling constant, $q=p_1+p_2=p_1^{\prime}+p_2^{\prime}$ is the momentum transfer, $T^{\alpha}=\frac{1}{2}\sigma^{\alpha}, \alpha=1,2,3$, are the generators for the Lie algebra $su(2)$ of the gauge group $SU(2)$ and $D_{m_Z\mu\nu}$ is the vector boson propagator given by
\[ D_{m_Z\mu\nu}(q)=D_{2,m_Z,\mu\nu}(q)=\frac{-\eta_{\mu\nu}+q_{\mu}q_{\nu}/m_Z^2}{q^2-m_Z^2}. \]
(In subsequent work we will determine the effect of using the correct Glashow Weinberg Salam (GWS)  model for the theory of the weak interaction.)
Thus
\begin{align*}
{\mathcal M}=&{\mathcal M}_{\alpha_1^{\prime}\alpha_2^{\prime}i_1^{\prime}i_2^{\prime}\alpha_1\alpha_2i_1i_2}(p_1^{\prime},p_2^{\prime},p_1,p_2)\\
=&-g_W^2\delta_{\alpha\alpha^{\prime}}T^{\alpha}_{i_1i_2}T^{\alpha^{\prime}}_{i_1^{\prime}i_2^{\prime}}\overline{v}(p_1,\alpha_1)\gamma^{\mu}u(p_2,\alpha_2)D_{m_Z\mu\nu}(q)\overline{u}(p_2^{\prime},\alpha_2^{\prime})\gamma^{\nu}v(p_1^{\prime},\alpha_1^{\prime}).
\end{align*}

Now consider
\[ X=\overline{v}(p_1,\alpha_1)\gamma^{\mu}u(p_2,\alpha_2)q_{\mu}q_{\nu}\overline{u}(p_2^{\prime},\alpha_2^{\prime})\gamma^{\nu}v(p_1^{\prime},\alpha_1^{\prime}). \]
Then
\begin{align*}
X&=\overline{v}(p_1,\alpha_1)\gamma^{\mu}u(p_2,\alpha_2)(p_{1\mu}+p_{2\mu})(p_{1\nu}^{\prime}+p_{2\nu}^{\prime})\overline{u}(p_2^{\prime},\alpha_2^{\prime})\gamma^{\nu}v(p_1^{\prime},\alpha_1^{\prime})\\
&=\overline{v}(p_1,\alpha_1)\gamma^{\mu}u(p_2,\alpha_2)(p_{1\mu}p_{1\nu}^{\prime}+p_{2\mu}p_{1\nu}^{\prime}+p_{1\mu}p_{2\nu}^{\prime}+p_{2\mu}p_{2\nu}^{\prime})\\
&\overline{u}(p_2^{\prime},\alpha_2^{\prime})\gamma^{\nu}v(p_1^{\prime},\alpha_1^{\prime})\\
&=\overline{v}(p_1,\alpha_1){\slas p}_1u(p_2,\alpha_2)\overline{u}(p_2^{\prime},\alpha_2^{\prime}){\slas p}_1^{\prime}v(p_1^{\prime},\alpha_1^{\prime})+\overline{v}(p_1,\alpha_1){\slas p}_2u(p_2,\alpha_2)\overline{u}(p_2^{\prime},\alpha_2^{\prime}){\slas p}_1^{\prime}v(p_1^{\prime},\alpha_1^{\prime})+\\
&\overline{v}(p_1,\alpha_1){\slas p}_1u(p_2,\alpha_2)\overline{u}(p_2^{\prime},\alpha_2^{\prime}){\slas p}_2^{\prime}v(p_1^{\prime},\alpha_1^{\prime})+\overline{v}(p_1,\alpha_1){\slas p}_2u(p_2,\alpha_2)\overline{u}(p_2^{\prime},\alpha_2^{\prime}){\slas p}_2^{\prime}v(p_1^{\prime},\alpha_1^{\prime})\\
&=\overline{v}(p_1,\alpha_1)(-m_e)u(p_2,\alpha_2)\overline{u}(p_2^{\prime},\alpha_2^{\prime})(-m_{\mu})v(p_1^{\prime},\alpha_1^{\prime})+\\
&\overline{v}(p_1,\alpha_1)(m_e)u(p_2,\alpha_2)\overline{u}(p_2^{\prime},\alpha_2^{\prime})(-m_{\mu})v(p_1^{\prime},\alpha_1^{\prime})+\\
&\overline{v}(p_1,\alpha_1)(-m_e)u(p_2,\alpha_2)\overline{u}(p_2^{\prime},\alpha_2^{\prime})(m_{\mu})v(p_1^{\prime},\alpha_1^{\prime})+\\
&\overline{v}(p_1,\alpha_1)(m_e)u(p_2,\alpha_2)\overline{u}(p_2^{\prime},\alpha_2^{\prime})(m_{\mu})v(p_1^{\prime},\alpha_1^{\prime})\\
&=0.
\end{align*}

Therefore
\[ {\mathcal M}=g_W^2(q^2-M^2)^{-1}\delta_{\alpha\alpha^{\prime}}T^{\alpha}_{i_1i_2}T^{\alpha^{\prime}}_{i_1^{\prime}i_2^{\prime}}\overline{v}(p_1,\alpha_1)\gamma^{\mu}u(p_2,\alpha_2)\eta_{\mu\nu}\overline{u}(p_2^{\prime},\alpha_2^{\prime})\gamma^{\nu}v(p_1^{\prime},\alpha_1^{\prime}), \]
where $M$ is the unknown mass of the Z$^0$ boson. We are taking the electron mass as known (i.e. as the reference mass). Also we can, from the result of Section~\ref{section:charged_leptons}, take the mass $m_{\mu}$ as known. Thus the only unknown mass in $\Phi_{\mathcal M}$ the mass $M(=m_Z)$ of the Z$^0$ boson. Hence
\begin{align*}
\Phi=\Phi(M,p_1^{\prime},p_2^{\prime},p_1,p_2)&=c(q^2-M^2)^{-2}\Psi,
\end{align*}
where $c$ is the constant given by
\[ c=\frac{1}{12}g_w^4\sum_{\alpha=1}^3\sum_{i_1,i_2,i_1^{\prime},i_2^{\prime}=1}^2|T^{\alpha}_{i_1i_2}T^{\alpha}_{i_1^{\prime}i_2^{\prime}}|^2. \]
and
\begin{equation} \label{eq:Z1}
\Psi=\frac{1}{4}\sum_{\alpha_1,\alpha_2\alpha_1^{\prime},\alpha_2^{\prime}=1}^2|\overline{v}(p_1,\alpha_1)\gamma^{\mu}u(p_2,\alpha_2)\eta_{\mu\nu}\overline{u}(p_2^{\prime},\alpha_2^{\prime})\gamma^{\nu}v(p_1^{\prime},\alpha_1^{\prime})|^2.
\end{equation}

Using the results of Section~\ref{section:charged_leptons} we have that 
\[ \Psi=8((p_1\cdot p_1^{\prime})(p_2\cdot p_2^{\prime})+(p_2\cdot p_1^{\prime})(p_1\cdot p_2^{\prime})+m_e^2(p_1^{\prime}\cdot p_2^{\prime})+m_{\mu}^2(p_1\cdot p_2)+2m_e^2m_{\mu}^2). \]

\begin{figure} 
\centering
\includegraphics[width=15cm]{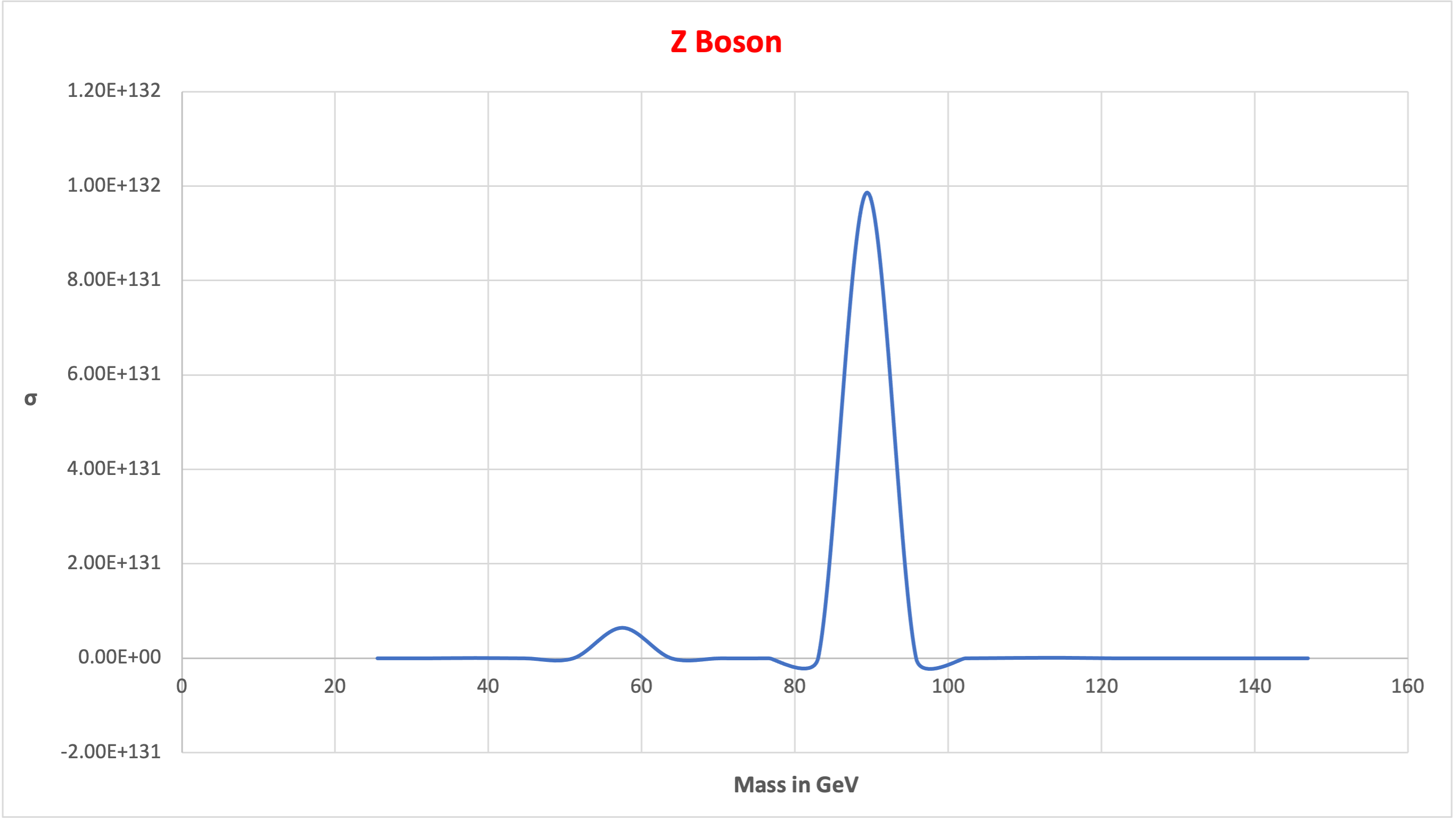}
\caption{Integral mass spectrum for the Z$^0$ boson \label{fig:Z_boson}}
\end{figure}

Plugging this form for $\Phi_{\mathcal M}$ into the equation for $\sigma_{{\mathcal M}}$ given in Section~\ref{section:fundamental_particle_masses} one can compute the integral mass spectrum for the family of gauge bosons associated with ${\mathcal M}$. This computation took less than a day. The parameters input to the program were as follows.
\begin{verbatim}
const double pi = 4.0 * atan(1.0);
const double m_e = 0.51099895; //electron mass in MeV 
const double m_mu = 105.7; // muon mass in MeV
double alpha_W = 1.0e-6;
double g_W = sqrt(4.0 * pi * alpha_W);
double Tiny = 1.0e-9;
double Start = 50.0e3*m_e;
double End = 300.0e3*m_e; 
int N_m = 20;
double delta_m = (End - Start) / N_m;
int N_int = 10;
double Lambda_int = End;
double delta_int = Lambda_int / N_int;
int N_int_angle = 10;

\end{verbatim}
The C++ code for carrying out this calculation can be found in Appendix C contained in the Supplementary Material for this paper as published in AIP Advances \cite{AIPAdv}. The resulting graph shown in Figure~\ref{fig:Z_boson} has a large peak located at 89.425 GeV which is 1.9\% percent deviation from the measured mass of $m_{\mbox{Z$^0$} }$= 91.1876 GeV. (Further investigation will need to be carried out in order to determine whether the small secondary peak would be removed by further computation.)  The values of $\alpha_W$ and hence $g_W$ are in fact not relevant in determining the Z boson mass because the effect of changing their values would be to simply multiply the $\sigma_{\mathcal M}$ function by a constant which would not affect the location of the peak. 

\subsection{Proposal for the determination of the masses of the W$^{\pm}$ particles and the neutrinos}

\begin{figure} 
\centering
\includegraphics[width=10cm]{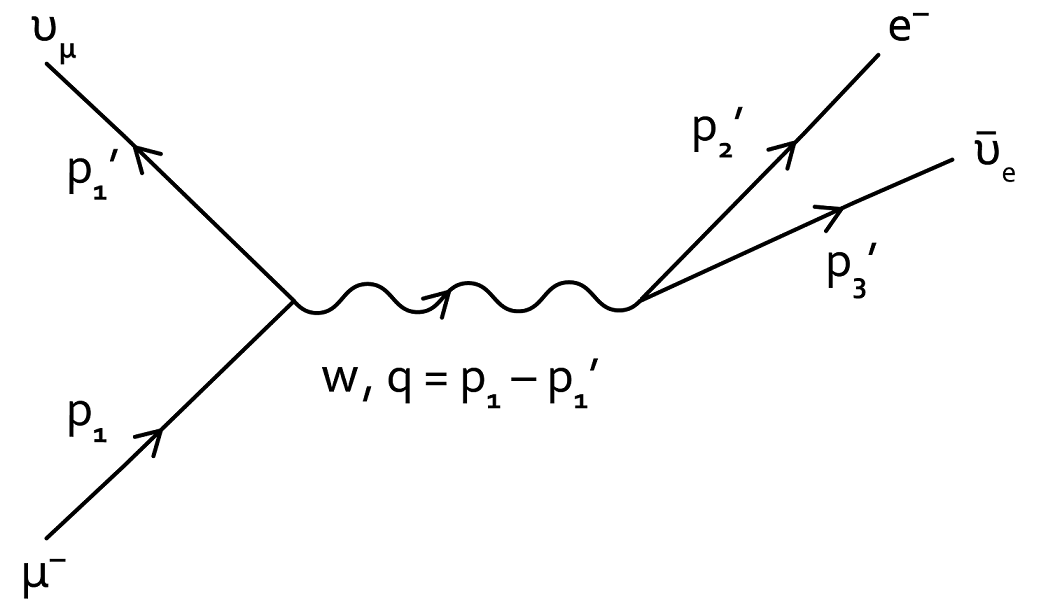}
\caption{Feynman diagram for muon decay\label{fig:muon_decay_diagram}}
\end{figure}

\begin{figure} 
\centering
\includegraphics[width=10cm]{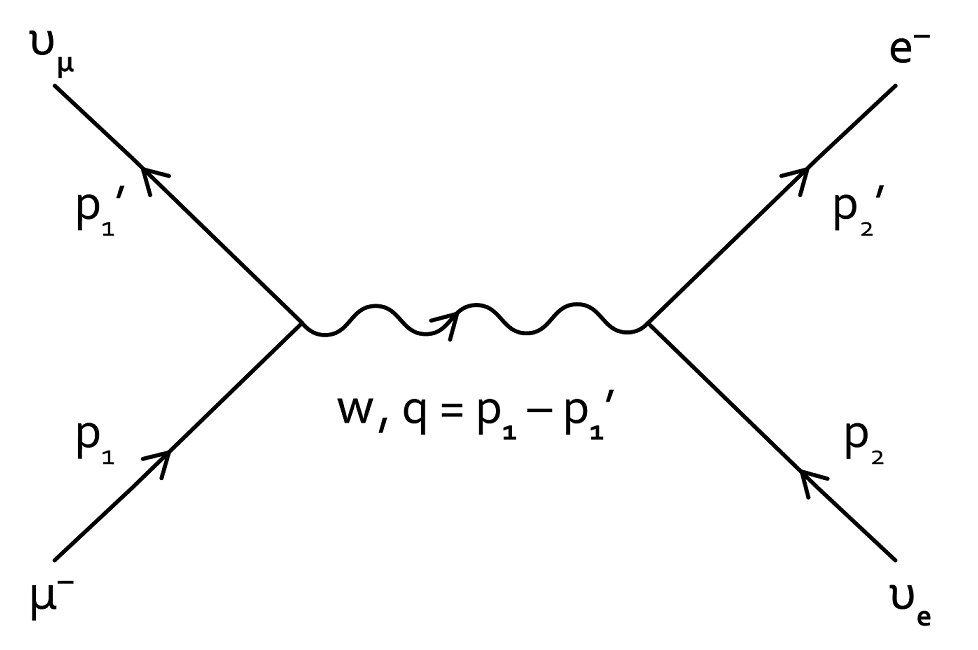}
\caption{Feynman diagram equivalent by crossing symmetry to diagram  for muon decay\label{fig:crossing_symmetry_diagram}}
\end{figure}

Consider muon decay $\mu^{-}\rightarrow e^{-}+\overline{\nu}_e+\nu_{\mu}$ as represented by the Feynman diagram of Figure~\ref{fig:muon_decay_diagram}.
By crossing symmetry this is equivalent to the process $\mu^{-}+\nu_e\rightarrow e^{-}+\nu_{\mu}$ whose Feynman diagram is shown in Figure~\ref{fig:crossing_symmetry_diagram}.  The Feynman amplitude for this process is
\begin{align*}
{\mathcal M} & = -g_W^2[\overline{u}(\nu_{\mu},p_1^{\prime},\alpha_1^{\prime})\gamma^{\mu}(1-\gamma^5)v(\mu^{-},p_1,\alpha_1)]D_{2,m_W,\mu\nu}(q) \\
    &  [\overline{v}(e^{-},p_2^{\prime},\alpha_2^{\prime})\gamma^{\nu}(1-\gamma^5)u(\nu_e,p_2,\alpha_2)], 
\end{align*}
where $q=p_1-p_1^{\prime}$ is the momentum transfer, $g_W$ is the weak force coupling constant, $u(\mbox{type},p,\alpha)$ and $v(\mbox{type},p,\alpha)$ are Dirac spinors for particles of type type, momentum $p\in H_m$ and polarizations $\alpha\in\{1,2\}$ and $D_{m_W,\mu_1\mu_2}$ is the W boson propagator.

The vertex for a process involving a W$^{\pm}$ boson is the classical V-A vertex. We will in subsequent work consider the possibility of using a GWS vertex.
The calculation needs to take into account the fact that the constraints of the problem now differ from those described by Eq.~\ref{eq:constraints}.

From the work of the previous sections we can take $m_e$ and $m_{\mu}$ as known.
Therefore there are three unknown masses in $\Phi_{\mathcal M}$, that is $m_W$, $m_{\nu_e}$ and $m_{\nu_{\mu}}$. 

The parameter $g_W$ is associated with a constant multiple of $\sigma_{{\mathcal M}}$ and therefore is not relevant to determining the masses $M_W,m_{\nu_e}$ and $m_{\nu_{\mu}}$. $\sigma_{{\mathcal M}}$ will be a function of these three unknowns on the three dimensional space $S=(0,\infty)^3$. One can compute $\sigma_{{\mathcal M}}$ and search for peaks over $S$. From these peaks one can determine the mass $m_W$ of the W$^{\pm}$ boson  and the neutrino masses.

\subsection{Proposal for the determination of the quark masses}

Consider the process $e^{+}e^{-}\rightarrow\overline{q}q$ at tree level. This process can occur through two routes (Schwartz, 2018, \cite{Schwartz}, p. 513), an intermediate photon or an intermediate Z$^0$ particle (there is no path involving an intermediate gluon because gluons carry color charge while $e^{+}$ and $e^{-}$ do not carry color charge).
The Feynman amplitude ${\mathcal M}$ for this process is
\[ {\mathcal M}={\mathcal M}_{\gamma}+{\mathcal M}_{\mbox{Z$^0$}}. \]
Thus we have that, up to color factors
\begin{align*}
&{\mathcal M}={\mathcal M}_{\alpha_1^{\prime}\alpha_2^{\prime}i_1^{\prime}i_2^{\prime}\alpha_1\alpha_2i_1i_2}(p_1^{\prime},p_2^{\prime},p_1,p_2),\\
&{\mathcal M}_{\gamma}={\mathcal M}_{\gamma,\alpha_1^{\prime}\alpha_2^{\prime}\alpha_1\alpha_2}(p_1^{\prime},p_2^{\prime},p_1,p_2),\\
&{\mathcal M}_{\mbox{Z$^0$}}={\mathcal M}_{\mbox{Z$^0$},\alpha_1^{\prime}\alpha_2^{\prime}i_1^{\prime}i_2^{\prime}\alpha_1\alpha_2i_1i_2}(p_1^{\prime},p_2^{\prime},p_1,p_2),
\end{align*} 
where
\begin{align*}
i{\mathcal M}_{\gamma}=&\overline{v}(p_1,\alpha_1)ie\gamma^{\mu}u(p_2,\alpha_2)iD_{\mu\nu}(p_1+p_2)\overline{u}(p_2^{\prime},\alpha_2^{\prime})ie\gamma^{\nu}v(p_1^{\prime},\alpha_1^{\prime}),
\end{align*}
and
\begin{align*}
i{\mathcal M}_{\mbox{Z$^0$}}=&\overline{v}(p_1,\alpha_1)ig_W\gamma^{\mu}T^{\alpha}_{i_1i_2}u(p_2,\alpha_2)i\delta_{\alpha\alpha^{\prime}}D_{m_Z\mu\nu}(q)\overline{u}(p_2^{\prime},\alpha_2^{\prime})ig_W\gamma^{\nu}T^{\alpha^{\prime}}_{i_1^{\prime}i_2^{\prime}}v(p_1^{\prime},\alpha_1^{\prime}),
\end{align*}

We are taking the mass of the electron as a given reference mass and we have computed the mass of the Z$^0$ boson. The only unknowns are the masses  $m_q$ which are what we want to compute.

It is anticipated that correctly computing color factors along the lines of the computations in (Schwartz, 2018, \cite{Schwartz} p. 514) and applying the machinery described in Section~\ref{section:description_of_method} will enable one to compute the integral mass spectrum for ${\mathcal M}$ and hence the quark masses. 

The issue with this formulation is that the mentioned processes also output other charged fermions apart from quarks. This issue is completely resolved in the same subsequent paper of ours which resolves the other issues of the paper mentioned above \cite{AIPAdv}. 

\section{Conclusion}

We have described a mathematical framework within which the gauge groups $U(1),SU(2)$ and $SU3)$ can be derived from the geometric structure of locally conformally flat space-time. Locally conformally flat space-time can have arbitrarily complicated geometric and topological structure.

It seems, remarkably, that one can determine elementary particle masses using simple tree level computations involving objects manifesting exact Lorentz invariance. 

The computation of the masses proceeds by computing the integral mass spectrum of the family of particles associated with any given covariant Feynman amplitude, seeking peaks in the integral mass spectrum and then taking the locations of the peaks to to be the masses of the particles in the family. 

We have successfully applied the method that we propose in the electroweak sector to the cases of the charged leptons $\mu$ and $\tau$ and the Z$^0$ particle.


\section*{Data availability statement}

The data that support the findings of this study are available from the author upon reasonable request. They were generated using the C++ computer programs listed in Appendices B and C which can be found in the Supplementary Material for this paper as pblished in AIP Advances \cite{AIPAdv}.

\section*{Appendix A: Smooth dynamic change $((p_1,p_2)\mapsto\Lambda(p_1,p_2)):H_{m_1}\times H_{m_2}\rightarrow O(1,3)^{\uparrow+}$ to CM frame}

We want to find a Lorentz transformation $\Lambda(p_1,p_2)$ smoothly depending on its arguments $p_1\in H_{m_1}$ and $p_2\in H_{m_2}$, $m_1,m_2>0$, such  that $\pi(\Lambda(p_1,p_2)p_1)=-\pi(\Lambda(p_1,p_2)p_2),\forall(p_1,p_2)\in H_{m_1}\times H_{m_2}$, where $\pi:{\bf R}^4\rightarrow{\bf R}^3$ is the canonical projection defined by $\pi((p^0,{\vct p}))={\vct p}$.

The first step is to find, in general, a function $p\mapsto\Xi(p)$ from $C_0=\{p\in{\bf R}^4:p^2>0,p^0>0\}$ to $K$ with the property that
\[ \Xi(p)p=(\zeta(p),{\vct0}),\forall p\in C_0, \]
where $\zeta:C_0\rightarrow(0,\infty)$ is given by $\zeta(p)=(p^2)^{\frac{1}{2}}$.

Once we have done that we can, given $p_1\in H_{m_1},p_2\in H_{m_2}$, compute $\Lambda(p_1,p_2)=\Xi(p_1+p_2)$ ($K$ acts on ${\bf R}^4$ by Lorentz transformations). Then $\Lambda(p_1,p_2)(p_1+p_2)=(\zeta(p_1+p_2),{\vct0})$ and the required result clearly follows. 

\subsection*{Determination of the function $(p\mapsto \Xi(p)):C_0\rightarrow K$}

We would like to determine a function $\Xi:C_0\rightarrow K$ such that
\begin{equation}\nonumber
\Xi(p)p=(\zeta(p),{\vct 0}),
\end{equation}
Identifying ${\bf R}^4$ with $iu(2)=\{\mbox{ hermitian\mbox{ $2\times2$ }matrices }\}$ in the standard way $K$ acts on ${\bf R}^4$ according to \cite{AMP}
\begin{equation}\nonumber
\kappa(M)=aMa^{\dagger},
\end{equation}
for
\begin{equation}\nonumber
\kappa=\left(\begin{array}{cc}
a & 0 \\
0 & a^{\dagger-1}
\end{array}\right)\in K.
\end{equation}
$(\zeta(p),{\vct 0})$ corresponds to the element $\zeta(p)=\zeta(p)1_2\in iu(2)$. Therefore we wish to find an $a=a(p)\in GL(2,{\bf C})$ with $|\mbox{det}(a)|=1$ such that $aM(p)a^{\dagger}=\zeta(p)$. This is equivalent to finding $a\in GL(2,{\bf C})$ with $|\mbox{det}(a)|=1$ such that $a^{\dagger}a=\zeta(p)M(p)^{-1}$.

Now since $\mbox{det}(M(p))=\zeta(p)^2$, if we find an $a\in GL(2,{\bf C})$ such that $a^{\dagger}a=\zeta(p) M(p)^{-1}$ then it must satisfy $|\mbox{det}(a)|=1$ since $|\det(a)|^2=\mbox{det}(a^{\dagger}a)=\mbox{det}(\zeta(p)M(p)^{-1})=\zeta(p)^2\mbox{det}(M(p))^{-1}=1$.

\begin{theorem} \label{theorem:Appendix_1_2}
Let $p\in C_0=\{p\in{\bf R}^4:p^2>0,p^0>0\}$. Then the eigenvalues of $M(p)$ are real positive.
\end{theorem}
{\bf Proof}
$M(p)$ is hermitian so its eigenvalues are real. We have
\begin{align*}
&\mbox{$\lambda$ is an eigenvalue of $M(p)$}\\
&\Leftrightarrow\mbox{det}(M(p)-\lambda)=0\\
&\Leftrightarrow\mbox{det}\left(\begin{array}{cc}
p^0+p^3-\lambda&p^1-ip^2\\
p^1+ip^2&p^0-p^3-\lambda
\end{array}\right)=0\\
&\Leftrightarrow(p^0-\lambda)^2-(p^1)^2-(p^2)^2-(p^3)^2=0\\
&\Leftrightarrow|\lambda-p^0|=|{\vct p}|\\
&\Leftrightarrow\lambda=p^0\pm|{\vct p}|.
\end{align*}
But $p\in C_0$ and so $p^0>|{\vct p}|$, from which the result follows. $\Box$.

Therefore the eigenvalues of $\zeta(p)M(p)^{-1}$ are real positive. Now diagonalize $\zeta(p)M(p)^{-1}$ so that
\begin{equation}\nonumber
\zeta(p)M(p)^{-1}=U(p)\mbox{diag}(\lambda_1(p),\lambda_2(p))U(p)^{-1},
\end{equation}
where $U(p)\in U(2)$ and $\lambda_1(p),\lambda_2(p)>0$. Then taking
\begin{equation*}
a(p)=U(p)\mbox{diag}(\lambda_1(p)^{\frac{1}{2}},\lambda_2(p)^{\frac{1}{2}})U(p)^{-1},
\end{equation*}
solves our problem. In fact, we take $\Xi:C_0\rightarrow K$ to be defined by
\[ \Xi(p)=\mbox{diag}(a(p),a(p)^{\dagger-1}). \] 

\subsection*{Lorentz invariance of the function $((p_1,p_2)\mapsto E(p_1,p_2))$}

\[ E(p_1,p_2)=\frac{1}{2}((\Lambda(p_1,p_2)p_1)^0+(\Lambda(p_1,p_2)p_2)^0)=\frac{1}{2}(\Lambda(p_1+p_2)(p_1+p_2))^0=\frac{1}{2}E_{\mbox{CM}}(p_1,p_2), \]
where
\[ E_{\mbox{CM}}(p_1,p_2)=\zeta(p_1+p_2). \]
$E_{\mbox{CM}}$ is called the the total CM energy of the process and is Lorentz invariant since
\[ E_{\mbox{CM}}(\Lambda p_1,\Lambda p_2)=\zeta(\Lambda p_1+\Lambda p_2)=\zeta(\Lambda(p_1+p_2))=\zeta(p_1+p_2)=E_{\mbox{CM}}(p_1,p_2),\forall\Lambda\in O(1,3)^{\uparrow+}. \]

\end{document}